# Role of Data-driven Regional Growth Model in Shaping Brain Folding Patterns

Jixin Hou[1], Zhengwang Wu[2], Xianyan Chen[3], Li Wang[2], Dajiang Zhu[4], Tianming Liu[5], Gang Li[2*], Xianqiao Wang[1]*

[1]School of Environmental, Civil, Agricultural and Mechanical Engineering, College of Engineering, University of Georgia, Athens, GA 30602, USA

[2]Department of Radiology and Biomedical Research Imaging Center, The University of North Carolina at Chapel Hill, NC 27599, USA

[3]Department of Epidemiology and Biostatistics, College of Public Health, University of Georgia, Athens, GA 30602, USA

[4]Department of Computer Science and Engineering, The University of Texas at Arlington, Arlington, TX 76019, USA

[5]School of Computing, The University of Georgia, Athens, GA 30602, USA

*Corresponding Author: gang_li@med.unc.edu, xqwang@uga.edu

## Abstract

The surface morphology of the developing mammalian brain is crucial for understanding brain function and dysfunction. Computational modeling offers valuable insights into the underlying mechanisms for early brain folding. Recent findings indicate significant regional variations in brain tissue growth, while the role of these variations in cortical development remains unclear. In this study, we unprecedentedly explored how regional cortical growth affects brain folding patterns using computational simulation. We first developed growth models for typical cortical regions using machine learning (ML)-assisted symbolic regression, based on longitudinal real surface expansion and cortical thickness data from prenatal and infant brains derived from over 1,000 MRI scans of 735 pediatric subjects with ages ranging from 29 post-menstrual weeks to 24 months.

These models were subsequently integrated into computational software to simulate cortical development with anatomically realistic geometric models. We comprehensively quantified the resulting folding patterns using multiple metrics such as mean curvature, sulcal depth, and gyrification index. Our results demonstrate that regional growth models generate complex brain folding patterns that more closely match actual brains structures, both quantitatively and qualitatively, compared to conventional uniform growth models. Growth magnitude plays a dominant role in shaping folding patterns, while growth trajectory has a minor influence. Moreover, multi-region models better capture the intricacies of brain folding than single-region models. Our results underscore the necessity and importance of incorporating regional growth heterogeneity into brain folding simulations, which could enhance early diagnosis and treatment of cortical malformations and neurodevelopmental disorders such as cerebral palsy and autism.

**Keywords**. cortical folding pattern; heterogeneous growth; regional growth model; symbolic regression; computational modeling

## 1. Introduction

As the central regulator of physiological activities, the human brain exhibits an intricate tissue structure marked by pronounced heterogeneity. During early development (16-40 post-menstrual weeks), the brain undergoes considerable expansion in its volume and cortical surface area, accompanied by the noticeable observation of cortical folds, with the convex peak known as gyri and the concave valley as sulci. Typically, cortical folding begins with primary folds, which are highly conserved across individuals [1], and gradually progresses to more complex secondary and tertiary folds that exhibit significant individual variability [2]. Emerging evidence links abnormal cortical folding patterns to various neurodevelopmental disorders such as autism [3], epilepsy [4], and schizophrenia [5], as well as cortical malformations like lissencephaly, pachygyria, and polymicrogyria [6]. Consequently, a comprehensive understanding of the development of cortical folding is crucial for early detection and treatment of cognitive impairments and neurodevelopmental disorders.

The development of cortical folding involves highly complex spatiotemporal patterns [7] influenced by multiple physiological factors, such as cranial constraints [8], axon maturation [9], cerebrospinal fluid [10], genetic expression [11], etc. Though the precise mechanisms underlying cortical fold formation remains elusive, increasing studies suggest that mechanics play a crucial role in modulating this process [12]. Brain tissue development occurs within a complex physical environment comprising the skull, meninges, and cerebrospinal fluid. While external forces such as constraints from the skull and cerebrospinal fluid pressure are expected to affect cortical folding patterns, experimental observations indicate that the internal forces exerted within brain tissue are more dominate at the onset of folding [13]. In line with these findings, Essen [14] firstly proposed the "axon tension" theory, which suggests that axonal tension pulls gyral walls together, leading to formation of cortical folds. In contrast, Nie, et al. [15] proposed that pushing force from axonal growth could generate folding shapes similar to those observed in real cortical surface. However, despite experimental evidence of forces acting on axons, their magnitude and direction appear insufficient to pull or push brain tissues into folds [16]. Conversely, the differential growth theory, which has garnered more experimental support [17, 18], posits that the mismatch in growth ratios between two-layered brain structure triggers the bulking of cortical surface [19]. Specifically, the fast-growing outer layer (gray matter), under connectivity constraint of slow-growing inner layer (white matter), generates excessive compressive forces that initiate cortical folding. Therefore, it is essential to investigate how varying tissue growth influences the initiation and regulation of cortical folding patterns.

Computational simulation, such as finite element method (FEM), have emerged as a powerful tool for modeling the spatiotemporally complex development of cortical folding patterns, due to its superiority in reproducibility and cost-effectiveness compared to experimental and longitudinal imaging methods [20]. To replicate folding patterns, most studies have assumed uniform growth patterns such as isotropic growth and purely tangential growth across growing tissues [21-25].

While these approaches have greatly advanced our understanding of cortical folding mechanisms, their effectiveness diminishes when comparing the simulated patterns to actual brain structures, particularly in brain-wide simulations [26, 27]. This discrepancy arises because uniform growth assumptions conflict with the heterogeneous nature of growth as observed in experimental and imaging investigations [28, 29]. Recent studies, including our own study [30], have shown that the cortical surface can be parcellated into 18 distinct regions, each exhibiting significantly different area expansion ratios from the third trimester of pregnancy to the first two years post-birth, as illustrated in Figure 1. Similar parcellation based on heterogenous development of cortical thickness have also been reported [31]. These observations underscore the pronounced spatiotemporal heterogeneity in cortical tissue growth.

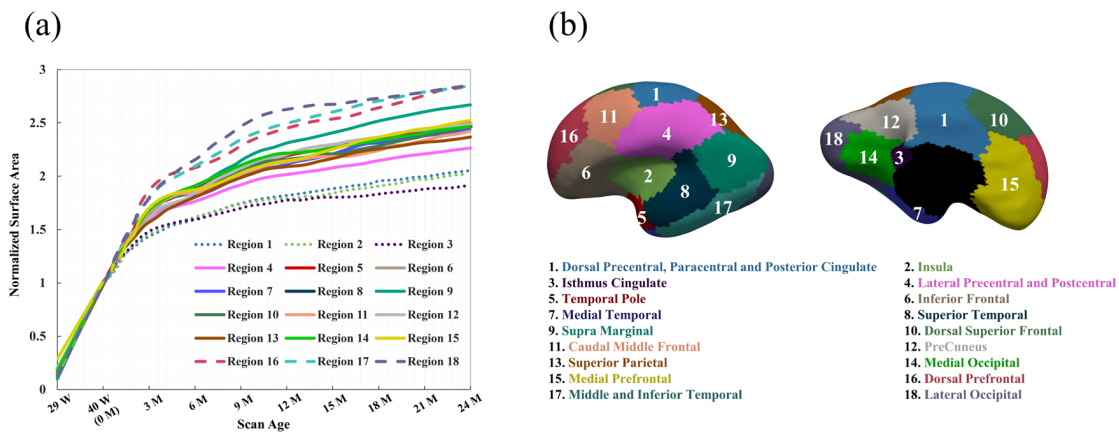

**Figure 1. Developmental trajectory of surface area in each region.** (a) Surface area is normalized based on the 40 post-menstrual weeks. W and M represent weeks and months, respectively; (b). Developmental regionalization map of 18 regions.

Despite its significance, the effect of growth heterogeneity on cortical folding patterns has been largely overlooked in computational studies. Budday, et al. [32] were among the first to investigate this effect by incorporating manually designed sinusoidal variations in growth distributions during folding simulations based on idealized models. Their findings revealed that regional variations in cortical growth primarily influence secondary instabilities, contributing to the irregularity of folding patterns. Wang, et al. [33] further explored this by simulating cortical folding evolution on theoretical brain models with regionally heterogeneous growth, linking the growth ratio to local curvature. Their results demonstrate the role of regional growth in shaping complex cortical folding patterns, characterized by fast-growing gyri and slow-growing sulci. Similar conclusions were drawn by Zhang, et al. [34]. While these studies have enhanced our understanding of the critical role of heterogeneous growth in cortical folding development, they have primarily focused on spatial heterogeneity and relied on idealized geometric models with manually defined growth patterns. These approaches deviate significantly from real brain scenarios. A comprehensive study examining the effects of realistic heterogeneous growth on cortical folding patterns in models that closely resemble real brains is still lacking.

In this study, we aim to explore how anatomically realistic regional growth affects the simulated cortical folding patterns on geometric models derived from real brains. We first employed ML-assisted symbolic regression to generate regional growth models using longitudinal surface area and cortical thickness data, including over 1,000 infant Magnetic Resonance Imaging (MRI) scans from developing prenatal and neonatal human brains. To our knowledge, this is the first study to directly characterize anatomically accurate growth models from such intensive real human brain images. These predicted growth models were then integrated into ABAQUS to simulate mechanical folding development using tangential differential growth theory, applied to regional geometric models constructed from initial brain surfaces. The spatiotemporal folding patterns were analyzed and compared with those observed in real brains, both qualitatively and quantitatively. Additionally, we conducted an in-depth analysis by decomposing the growth's value and trajectory to access their individual impacts on cortical folding patterns. Our results highlight the importance of considering growth heterogeneity in simulations of brain folding development. This approach could potentially contribute to early diagnosis of cortical malformations such as lissencephaly and polymicrogyria and improve treatments for neurodevelopmental disorders like epilepsy and autism.

## 2. Methods

### 2.1. Biomechanics in modeling brain folding

Differential growth theory indicates that cortical folding during brain development arises due to disparities in growth ratio between the faster-growing cortical layer and the slower-growing white matter layer. To investigate the effect of regional growth on folding evolution in the developing brain, we conducted nonlinear finite element simulations grounded in continuum mechanics theory and the finite growth assumption. In describing deformation kinematics, we introduce a one-to-one mapping denoted as $x = \varphi(X)$, which carries a material particle initially positioned at $X$ in the reference configuration $\mathcal{B}_0$ to its new position $x$ in the current configuration $\mathcal{B}_t$. Leveraging the theory of multiplicative decomposition proposed by Rodriguez, et al. [35], we decompose the deformation gradient $F$ and the Jacobian $J$ into elastic and growth components:

$$F = \nabla_X \varphi = F^e \cdot F^g \quad \text{and} \quad J = \det F = J^e \cdot J^g. \tag{1}$$

We assume that brain tissue growth follows an orthotropic manner,

$$F^g = g_t I + (g_r - g_t)\hat{n} \otimes \hat{n} \quad \text{and} \quad J^g = \det F^g, \tag{2}$$

where $g_t$ and $g_r$ are the relative growth ratio defined in the tangential (in-plane) and radial (out-of-plane) directions, respectively, of which the expressions ($g_t = g_t(t)$, $g_r = g_r(t)$) are derived from the symbolic regression predictions, as discussed in the following section. $g_t$ equating to $g_r$ indicates an isotropic growth scenario, while $g_r = 1$ denotes tangential growth. The vector $\hat{n}$ is the local surface normal, oriented positively in the out-of-plane direction. Growth kinematics alone generate a stress-free state, which must be balanced by the elastic deformation gradient introduced by material confinement to avoid incompatible deformation [36].

We characterized the constitutive behavior of the brain tissue through the following neo-Hookean strain energy function, parameterized exclusively in terms of the elastic components of deformation gradient $F^e$ and its Jacobian $J^e$,

$$\Psi(F^e) = \frac{1}{2}\lambda \ln^2(J^e) + \frac{1}{2}\mu(F^e : F^e - 2\ln J^e - 3), \tag{3}$$

where $\lambda$ and $\mu$ are the Lamé constant. In accordance with thermodynamic principles, the first Piola-Kirchhoff stress $\boldsymbol{P}$ works conjugate with the deformation gradient $\boldsymbol{F}$, particularly $\boldsymbol{F}^e$, since only the elastic part of deformation produces stress,

$$\boldsymbol{P} = \frac{\partial \Psi}{\partial \boldsymbol{F}} = \frac{\partial \Psi(\boldsymbol{F}^e)}{\partial \boldsymbol{F}^e} : \frac{\partial \boldsymbol{F}^e}{\partial \boldsymbol{F}} = \boldsymbol{P}^e \cdot \boldsymbol{F}^{g-T}. \tag{4}$$

In the absence of body force, the balance of linear momentum reduces to the vanishing divergence of the first Piola-Kirchhoff stress $\boldsymbol{P}$:

$$\text{Div } \boldsymbol{P} = 0 \tag{5}$$

**2.2. Symbolic regression for discovering growth models**

Inspired by Darwinian principles of natural selection, symbolic regression autonomously uncovers mathematical relationships exclusively from provided data without requiring prior knowledge, thereby significantly enhancing the interpretability and flexibility of the model discovery process [37]. It has demonstrated promising applications in model characterization [38-40] and parameter calibration [41]. Symbolic regression operates through a process known as genetic programming (GP). During GP execution, functional expressions are efficiently formatted using a binary-tree structure, which consists of nodes and branches.

A complete tree structure, as illustrated in Figure 2a, involves variables, mathematical operators (either unary or binary), and constants. The evolution process experiences the genetic operations of evaluation, selection, mutation, and crossover, while the latter two are essential for updating the tree structure. Specifically, the mutation operation amplifies population's genetic diversity by randomly altering some nodes in an expression tree, as exemplified in Figure 2b, where a new offspring is generated by substituting the exponential operator (exp) with the hyperbolic tangent (tanh). Conversely, the crossover operation allows the algorithm to create new offspring by combining building blocks from different parent candidates, as demonstrated in Figure 2c. This iterative process of evaluation, selection, mutation, and crossover continues until the optimal expression is obtained or the maximum number of generations is reached, whichever is reached first [37].

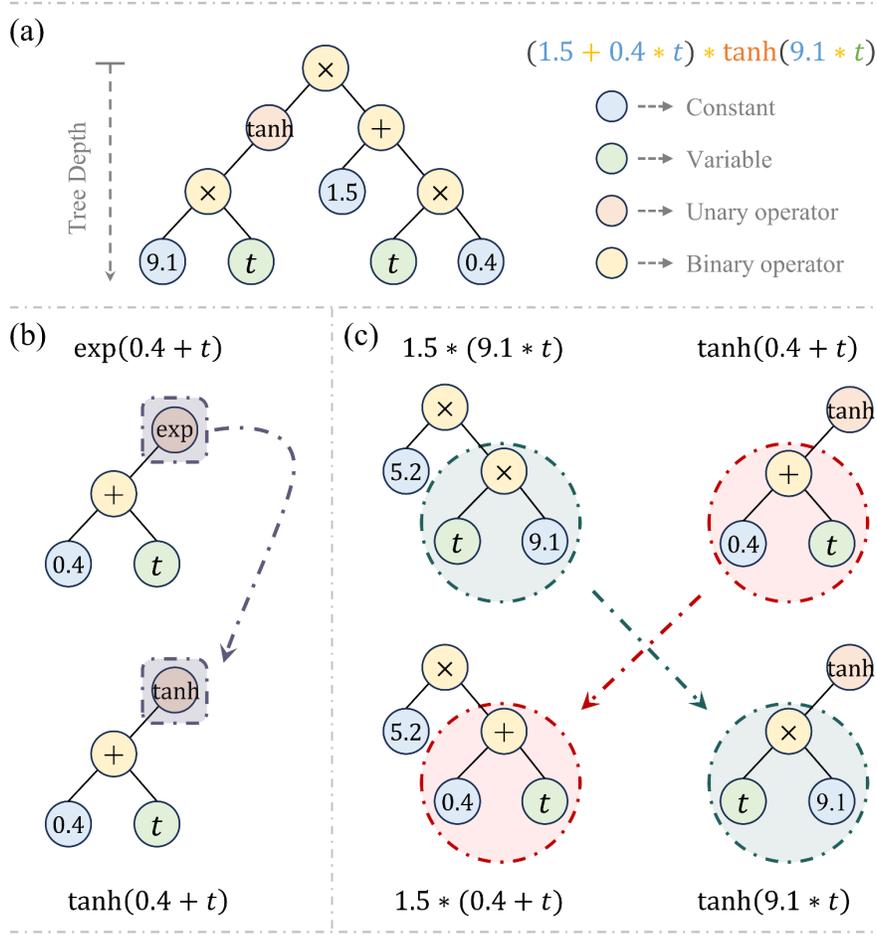

**Figure 2. Structure and operations of expression tree.** (a). Representation of tree structure for an algebraic expression; (b). An example of the mutation operation; (c). An example of the crossover operation between two parent trees.

In this study, we employed symbolic regression to discover appropriate growth models for the human brain cortex. The raw data includes the measured surface area of each parcellated region in the developing brain cortex along with the corresponding gestational ages, as shown in Figure 3a. To facilitate computational implementation, we firstly converted these data into unitless relative growth ratio $g(t)$ and virtual time $t$ using the following formulas:

$$g(t) = \sqrt{\frac{S(t)}{S_0}} \quad \text{and} \quad t = \frac{GA(t) - GA_0}{GA_{\max} - GA_{\min}}, \tag{6}$$

where $S_0$ denotes the surface area measured at the initial gestational age, $GA_0$ (29 post-menstrual weeks). $GA_{\max}$ and $GA_{\min}$ corresponds to the maximum and minimal gestational age within the measured data range, herein, their values are 29 post-menstrual weeks and 24 postnatal months of age, respectively. Through the above conversion, the value of $t$ ranges from 0 to 1, which serves as the input for the symbolic regression algorithm to find the optimal growth model $g(t)$ for each region, as illustrated in Figure 3b.

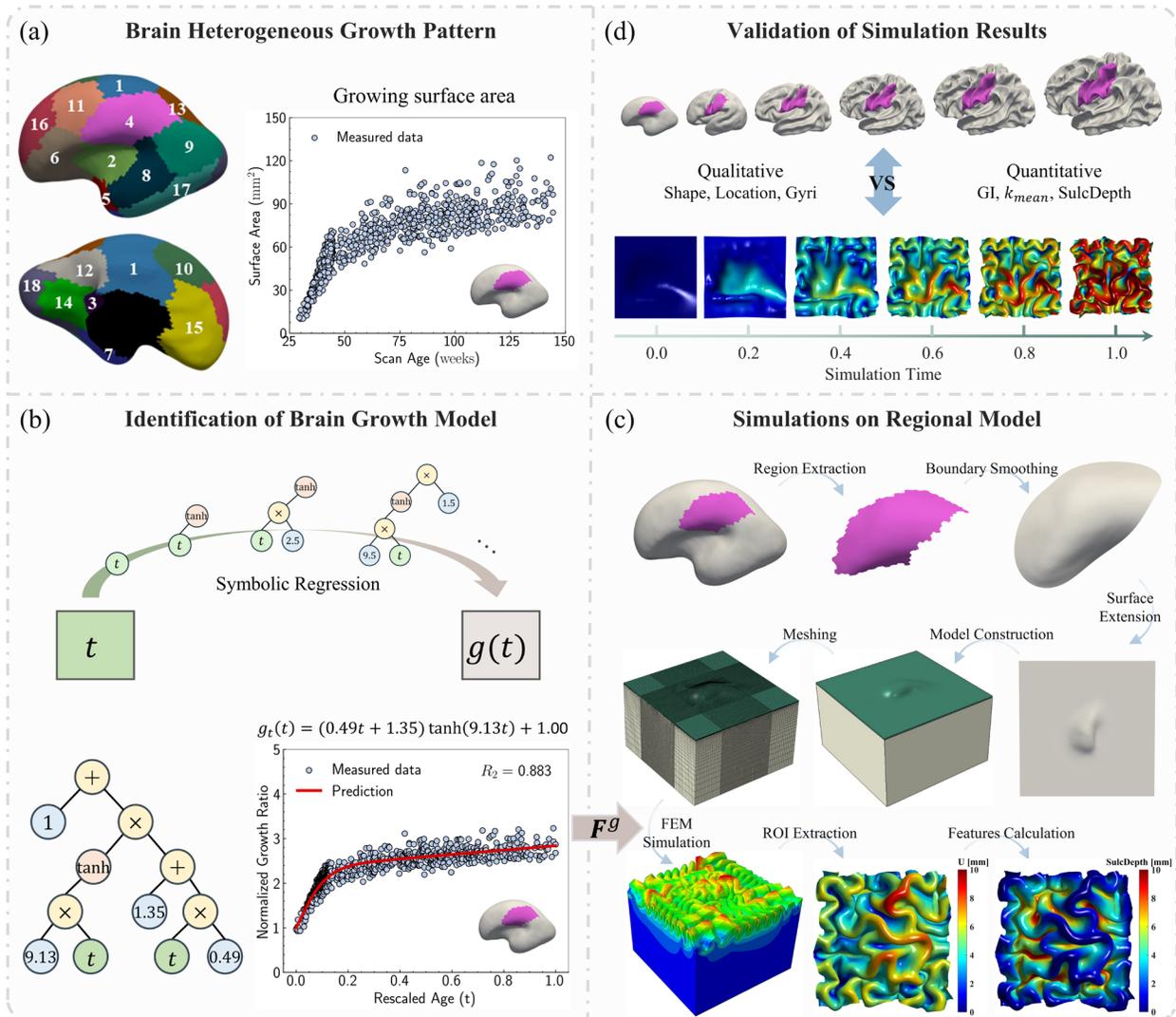

**Figure 3. Schematic diagram of this research.** (a) Cortical heterogeneous growth map with 18 regions determined based on the developmental patterns of surface area from 29 post-menstrual weeks to 24 postnatal months of age; (b). Symbolic regression algorithms used to identify the mathematical growth models for each region, exemplified with region 4; (c). Construction process of finite element model for each region, ROI means region of interest. (d). Qualitative and quantitative comparison between the simulation results and real human brains measures along the developmental timeline.

All symbolic regression analyses were performed using PYSR [42], a powerful open-source package developed alongside the Julia library *SymbolicRegression.jl*. During the training of symbolic regression, the binary operators are restricted to addition (+), multiplication (*), and polynomial ($t^n$), while the unary operators are limited to the exponential (exp), hyperbolic tangent (tanh), square ($t^2$), cube ($t^3$), and natural logarithmic (ln) functions. The training time is set to 30 minutes or a maximum of 1000 iterations, whichever is reached first. The maximum depth of the expression tree is set to 10, and the maximum complexity is constrained to 100. Notably, the nested behavior of the exponential, hyperbolic tangent, and logarithmic functions is forbidden, while the square and cube operators are restricted to occur, if desired, only once inside the exponential, hyperbolic tangent, and logarithmic functions. We adopt the default criterion ("best") to guide the model selection process. For each algorithm, the training is repeated at least three times, and favorable candidates are selected as the target models. All training was performed on a Legion PC equipped with a six-core Intel Core I7-8750H 2.2GHz CPU, 4 GB NVIDIA GTX 1050Ti GPU, and 24GB of memory.

### 2.3. Computational modeling of a developing brain

To simulate the folding evolution of the developing brain, we constructed a three-dimensional double-layer patch model based on the geometries of a human brain at 29 post-menstrual weeks, as depicted in Figure 3c. Initially, the regional brain inner surface (the interface between the gray matter and white matter) was extracted using the parcellation map provided by Huang, et al. [30]. After appropriate smoothing, the surface boundary was extended by 20-50 mm along its local curvature. This extension length varied depending on the local geometry and was chosen to be as large as possible without causing surface penetration. Subsequently, we interpolated the extended surface with a flat plane of 80 mm × 80 mm and merged these two surfaces using Boolean operations. The connecting areas were further smoothed to ensure a natural transition of the curvature. This extension and interpolation ensured that the model's dimensions were large enough compared to the wavelength of folded patterns observed in experiments, thereby preventing boundary effects [25]. Additionally, the squared boundaries significantly simplified the prescription of boundary conditions during modeling. We then shifted the interpolated surface upwards by 2 mm to form the initial cortical layer and extended the squared boundary downwards by 50 mm to generate the initial white matter layer. This design was based on experimental observations in neonatal human brains, which indicate that the cerebral cortex is a thin layer with a thickness of 2-3.5 mm, while the core has a much greater thickness of around 50 mm [43]. Consequently, the base model's dimensions were approximately 80 mm × 80 mm × 50 mm (excluding cortical thickness), as illustrated in Figure 4a, where $l = 80$ mm, $h_s = 50$ mm, and $h_c = 2$ mm. It should be noted that our objective was to investigate the effect of regional cortical growth on cortical folding evolutions, so the effect of axonal fibers was ignored in this study, despite their indispensable role in affecting cortical folding patterns [23, 44].

All simulations were performed using the commercial software ABAQUS (Dassault Systems, Paris, France) [45]. Dynamic-explicit solver was employed due to its superior performance in solving nonlinear, dynamic, and larger deformation problems [46, 47]. Both the gray and white matters were modeled as incompressible neo-Hookean materials, with elastic stiffness values of 0.31 kPa for the cortical and 0.45 kPa for the white matter layer [48]. Orthotropic growth was defined for the cortical layer, while isotropic growth was applied to the white matter layer. In our modeling approach, growth was simulated using thermal expansion, considering the analogy between the volumetric growth and the thermal expansion [49]. The expansion ratio $\alpha(t)$ correlates to the growth ratio as $\alpha(t) = g(t) - 1$. Specifically, for the white matter layer, the expansion ratio was defined as $\alpha(t) = 0$, while for the cortical layer, $\alpha(t) = g_r(t) - 1$ was applied to out-of-plane growth and $\alpha(t) = g_t(t) - 1$ to in-plane growth, as shown in Figure 4b. The customized growth models ($g_r(t)$ and $g_t(t)$), derived from symbolic regression, were implemented into the finite element algorithm through a user-defined subroutine *VUEXPAN*. Symmetric boundary conditions were prescribed on the four sides of the model and the bottom surface of the white matter layer was fixed. Free boundary conditions were applied to the top surface of the cortical layer, accompanied by a self-contact constraint to prevent self-penetration.

The total simulation time was set to 1 s. To avoid computational instabilities, the maximum time step was determined as $dt = 0.05a\sqrt{\rho/K}$, where $a$ is the average mesh size, $\rho$ is the mass density, $K$ is bulk modulus [50]. Temperature variation was applied using a sigmoidal smooth step function.

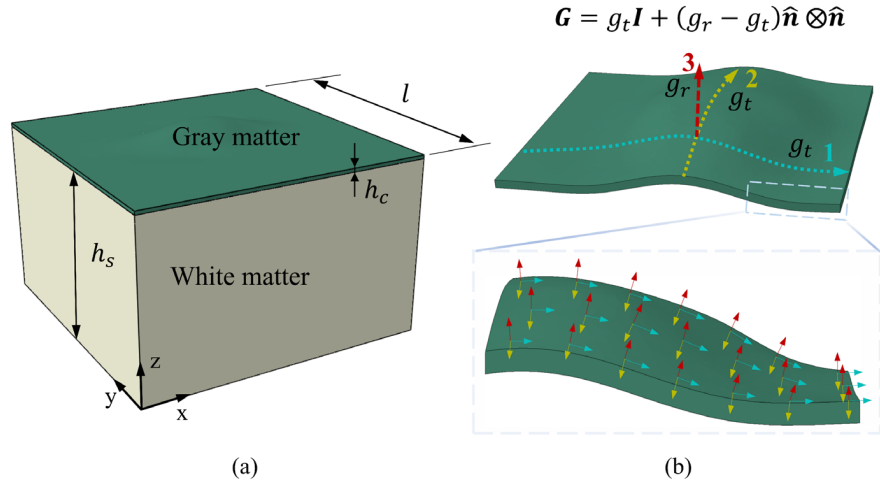

**Figure 4. Geometric model for modeling regional brain growth**. (a). Initial status of a FEM model, $h_c$ and $h_s$ represent the layer thickness of gray matter and white matter, respectively; (b). Orthogonal growth patterns, $g_t$ and $g_r$ indicate the tangential growth ratio within the plane and radial growth ratio in the thickness direction, respectively. Zoom-in shows material orientation assigned for the local region.

Structural meshing with the element type C3D8R was conducted for both the cortical and white matter layer, as shown in Figure 3c. To determine the appropriate mesh size, we conducted a mesh sensitivity analysis with mesh size ranging from 0.3 mm to 0.8 mm (Supplementary Material, Figures S1 and S2). Based on the mesh convergence analysis—where simulation results with the coarsest mesh closely matched those of the finest mesh—we selected a mesh size of 0.5 mm for all models. This results in 84,700 elements for the cortical layer and 278,300 elements for the white matter layer. During the simulations, we recorded the coordinates and displacements of the region of interest (ROI) for each frame, as shown in Figure 3d. The ROI was defined as the smallest square area encompassing the extracted brain surface region. Although this definition introduces some redundant areas, which potentially biases the quantitative measurements, it serves our primary goal: to compare the effectiveness of the regional growth model with the widely used isotropic growth theory. The inclusion of these redundant areas does not significantly impact this comparative analysis. Additionally, defining the ROI in this manner simplifies the partitioning process in ABAQUS and facilitates area reconstructions during postprocessing. All simulations were performed on a Dell workstation equipped with a 16-core Intel(R) Xeon(R) CPU E5–2687 W @ 3.1 GHz, and 64 GB of memory.

### 2.4. Postprocessing and quantitative metrics

After the simulations, the recorded coordinates and displacements were first extracted from the result file using Python and subsequently imported into MATLAB to reconstruct the deformed surface. During reconstruction, the surface was interpolated five times to generate a sufficiently smooth surface, and the original quadrilateral surface mesh was transformed into a triangular mesh, facilitating the calculation of quantitative features such as curvatures, gyrification index, and sulcal depth in MATLAB.

*Curvatures*: The curvature of a surface describes the degree to which it deviates from being flat at a given point. Normal curvature is defined as the inverse of the radius of the best-approximated curve from a surface normal slice in a given direction. Considering all directions, we obtain the curvature matrix, typically represented by the Weingarten matrix. Its principal decomposition gives the principal curvatures, which correspond to the maximum and minimum values of the surface's normal curvature in different directions ($k_1$ and $k_2$). The average of two principal curvatures denotes the mean curvature ($k_H = (k_1 + k_2)/2$, while the product of the principal curvatures yields the Gaussian curvature ($k_G = k_1 \cdot k_2$). In this study, we focused on the mean curvature due to its extensive application in brain cortical folding analysis [50-52].

The curvatures of the triangular mesh were calculated using the method introduced by Meyer, et al. [53], where finite volume discretization was employed to estimate the local integral of mean curvature over the normal areas of triangular faces associated with each point. If the surrounding faces are obtuse triangles, the barycentric area was calculated; otherwise, the Voronoi area was used. However, the calculated mean curvature is dependent on the geometry's shape or size, meaning its magnitude varies with brain scales. To address this, we further introduced a non-dimensional measure of mean curvature using the method provided by Balouchzadeh, et al. [54], where the mean curvature is multiplied by a characteristic length $l_c$,

$$l_c = \sqrt{A_t/4\pi} \quad \Rightarrow \quad k_H^* = k_H \cdot l_c, \tag{7}$$

where $A_t$ is the surface area. The dimensionless mean curvature was calculated for each point to provide a qualitative representation, while the absolute value of the dimensionless mean curvature was averaged across all model points for quantitative comparison. In the remainder of the manuscript, we use the term "curvature" to refer to "dimensionless mean curvature" for clarity.

*Gyrification index.* To quantitatively describe the folding complexity of the deformed brain surface, we introduced a global folding metric: the three-dimensional gyrification index (GI). The GI is defined as the ratio of the total cortical surface area to the area of convex hull that completely encloses the convoluted surface [8],

$$\text{GI} = \frac{area\ of\ coritcal\ surface}{area\ of\ convex\ hull}. \tag{8}$$

To calculate the GI, we first defined a fully enclosed convex hull comprising all points of the deformed cortical surface, then we discretized and filtered this surface to ensure it completely encloses the deformed surface with minimum surface area. Finally, we measured the area of discrete convex hull, which serves as the denominator in the GI calculation.

*Sulcal depth*: Sulcal depth (SulcDepth) is another quantitative measure capable of reflecting the extent of the folding in brain regions. Although Numerous methods have been suggested for computing sulcal depth [52, 55], a well-defined computation remains elusive. In this study, we adopted the approach introduced by Wang, et al. [50], calculating SulcDepth as the distance between the deformed mesh surface and its convex hull, which was previously defined in calculating the GI. Specifically, for each vertex on the deformed surface, we firstly determined its projection point on each discrete triangular surface of the convex hull. Subsequently, we computed the distance between the vertex and its corresponding projection point, with the shortest distance serving as the sulcal depth for that vertex. Given the convex nature of the enclosed hull, the shortest distance always exists between the vertex and a consistent piece of the convex hull, as demonstrated in the Result section. SulcDepth was calculated for each vertex for qualitative representation, and the values were averaged across all model points to ensure quantitative comparisons.

**2.5. Brain imaging data**

In this study, we aimed to conducted symbolic regression to identify tangential growth models by measuring brain surface area data from two high-quality, publicly available imaging datasets focused on early brain development: the dHCP and the BCP. The dHCP dataset includes 549 MRI scans from 500 healthy neonates (279 males, 221 females) aged between 29 and 45 post-menstrual weeks [56]. These MR images were obtained using a Philips 3T scanner with a 32-channel neonate-dedicated head coil at St. Thomas Hospital, London, UK [57]. The BCP dataset comprises 488 longitudinal MRI scans from 235 healthy, term-born infants (108 males, 127 females) scanned between 0.3 months and 24 months of age. These images were collected using 3T Siemens Prisma MRI scanners equipped with Siemens 32-channel head coils at The University of North Carolina at Chapel Hill and University of Minnesota. To investigate the developmental regionalization of cortical surface area expansion during pregnancy and infancy, a data-driven method known as nonnegative matrix factorization (NMF) was utilized to partition the cortical surface into distinct regions by clustering cortical vertices with similar developmental patterns. For more details, please refer to our recent work [30].

The radial growth model was characterized using thickness data measured in the developing neonatal brain [31]. It is important to note that the radial growth dataset consisted exclusively of postnatal measurement. Consequently, we could not perform symbolic regression on a continuous dataset spanning from 29 post-menstrual weeks to 24 postnatal months, as we did for the tangential growth model. To address this limitation, we assumed a linear increase in cortical thickness based on brain imaging analyses [58] and applied linear extrapolation for the perinatal period (29-40 post-menstrual weeks). Notably, we introduced several constraints to the symbolic regression algorithm: no growth occurs at the initial stage ($g_r(GA = 29) = 1$), and the derivatives at the connection stage were required to be smooth ($g'_r(GA = 40) \in C^0$).

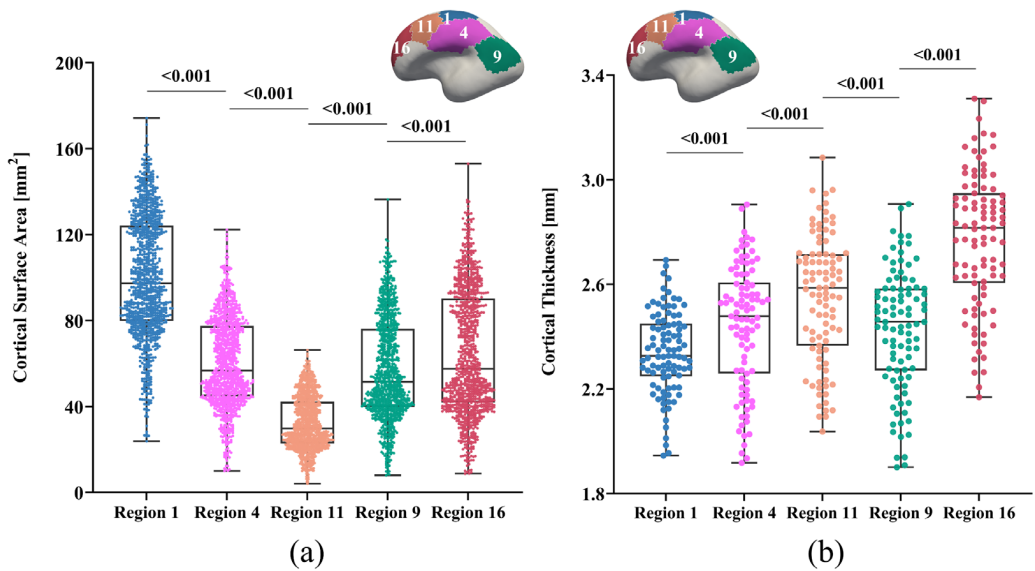

**Figure 5. Raw data of cortical surface area and thickness for five selected regions**. Analysis reveals significant differences in the development of surface area (a) and thickness (b) among these regions, with all *p*-values less than 0.001. Insets show spatial locations of regions on the cortical surface. Dots denote MRI scans.

We constructed all computational models using publicly available 4D infant cortical surface atlases [59, 60]. These atlases also served as the base model for measuring quantities such as curvature, GI, and SulcDepth, providing a baseline for quantitative comparison. In our previous study, the cortical surface was parcellated into 18 distinct regions based on the NMF method [30]. For this study, however, the primary objective was to compare the effectiveness of the regional growth model with classic growth theories. Therefore, we selected five representative regions that exhibit significant distinctions in surface area and cortical thickness. These regions are slow-growing Region 1, medium-growing Regions 4, 9, and 11, and fast-growing Region 16, corresponding to the cortical areas of "Dorsal Precentral, Paracentral and Posterior Cingulate", "Lateral Precentral and Postcentral", "Supra Marginal", "Caudal Middle Frontal", and "Dorsal Prefrontal", respectively. As illustrated in Figure 5, the development of surface area and cortical thickness shows significant differences among these regions, with the *p*-values all smaller than 0.001. Here, we conducted paired Student's t-test to compute the significance of these differences, with the null hypothesis being no difference in surface area or cortical thickness between the two regions.

## 3. Results

In this study, we first employed a symbolic regression algorithm to identify suitable growth models using longitudinal surface area and cortical thickness data from developing human brains, with ages ranging from 29 post-menstrual weeks to 24 postnatal months. These growth models were then integrated into ABAQUS via a user-defined subroutine to simulate mechanical folding evolutions caused by differential growth in double-layer structures. The dynamic folding patterns were analyzed and compared with anatomically realistic brain imaging models both in qualitative and quantitative manner. Subsequently, we substituted the regional growth model with classical growth models commonly used in previous studies, such as isotropic growth and purely tangential growth. We evaluated and compared the effectiveness of each model in modeling cortical folding development using quantitative metrics introduced in Section 2.4. We further investigated whether the accuracy of modeling is more influenced by the value of growth ratio or the trajectory of the growth model. Finally, we built a multi-regional model incorporating three regions with distinct growth patterns predicted by symbolic regression.

### 3.1. Regional growth models identified from symbolic regression

We investigated suitable tangential growth models through symbolic regression using the in-plane growth ratio data, normalized by the initial growth ratio as shown in Equation (6). This normalization ensures that the growth ratio starts from an initial value of 1. Moreover, given that the cortical surface area continues to increase beyond 24 postnatal months [61, 62], the derivative of the growth ratio at this point must remain positive ($g_t^t > 0$). These constraints were incorporated into the symbolic regression algorithm by customizing the loss function to impose a significant penalty for any violation. Figure 6 illustrates the tangential growth model identified for each representative region using symbolic regression. All models exhibit satisfactory accuracies, with

$R^2$ values exceeding 0.85, effectively capturing the growth patterns characterized by an initial linear increasing phase transitioning into a gradually steady phase, forming an upper-sigmoid shape. Intriguingly, among the predefined operators such as "exp", "tanh", "ln", and polynomials, as introduced in Section 2.2, the symbolic regression consistently selected a combination of tanh and polynomials. Moreover, a consistent model format, $g_t(t) = (a*t + b) * \tanh(c*t) + 1$, where $a$, $b$, and $c$ are constants, was discovered for almost all regions except for region 1. This finding suggests a promising model choice for charactering brain in-plane growth patterns.

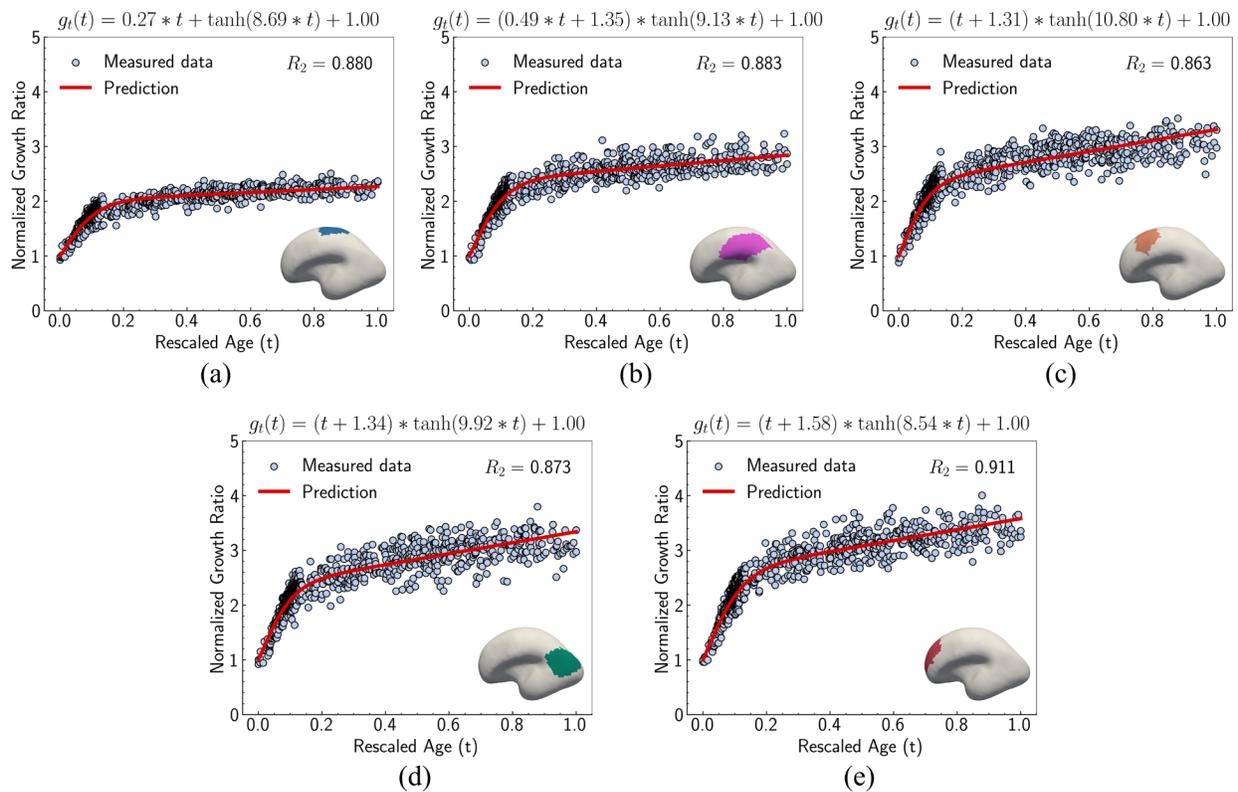

**Figure 6. Tangential growth models discovered for five selected regions.** In-plane growth predicted by symbolic regression algorithm for (a) region 1, (b) region 4, (c) region 11, (d) region 9, and (e) region 16, respectively. $R^2$ indicates the goodness of fit. Mathematical forms of growth models are presented at the top of each figure. Insets show spatial locations of regions on the cortical surface. Dots denote MRI scans.

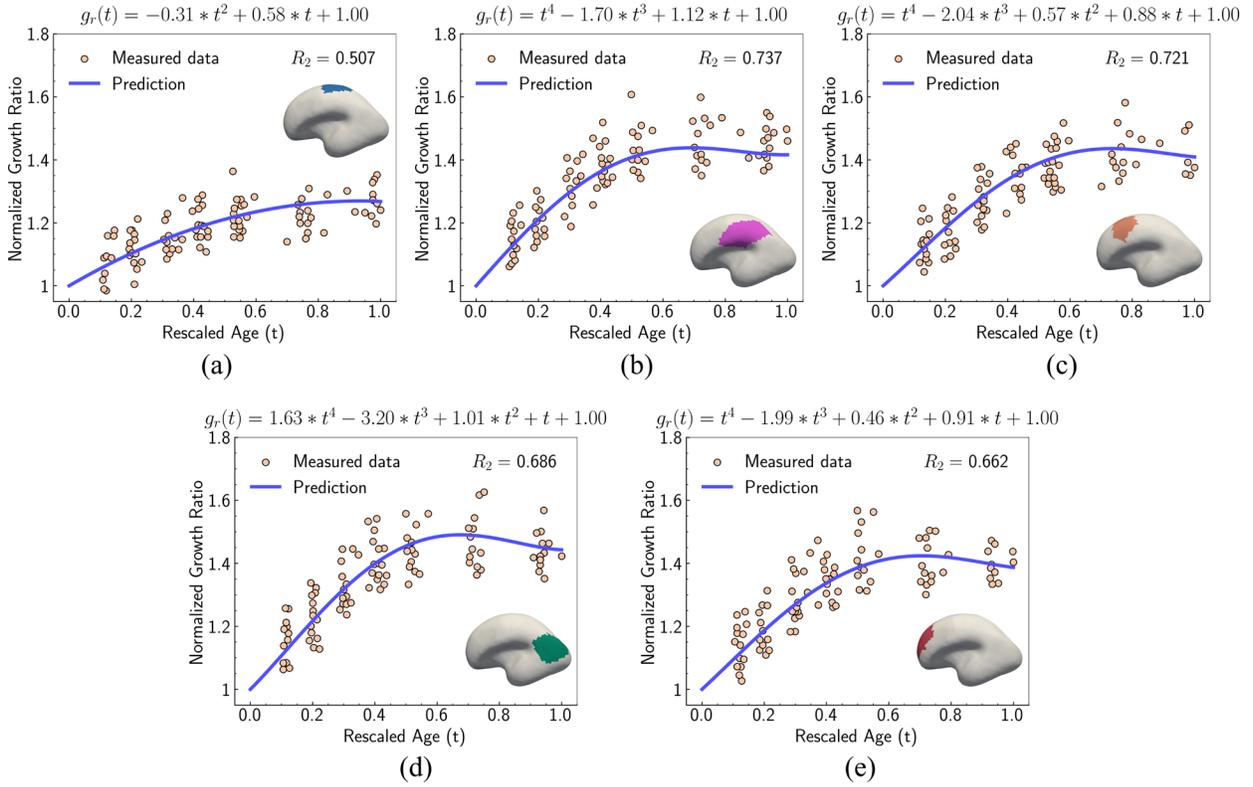

**Figure 7. Radial growth models discovered for five selected regions.** Growth of thickness predicted by symbolic regression algorithm for (a) region 1, (b) region 4, (c) region 11, (d) region 9, and (e) region 16, respectively. $R^2$ indicates the goodness of fit. Mathematical forms of growth models are presented at the top of each figure. Insets show spatial locations of regions on the cortical surface. Dots denote MRI scans.

For out-of-plane growth models, we applied distinct conditions to constrain growth behaviors. In addition to the initial value constraint ($g_r|_{t=0} = 1$), we incorporated an additional constraint to ensure the derivative is smooth at the connection stage ($g'_r|_{GA=40} \approx g'_r|_{t=0.1} \in C^0$) due to the use of linear extrapolation to address data scarcity during the perinatal period. Moreover, as the cortical thickness development tends to be stabilized within the first two years after birth [63, 64], the derivative at the phase end must be zero ($g'_r|_{t=1} = 0$). Figure 7 shows the radial growth models identified for each representative region through symbolic regression. Despite significant variance, the predicted growth model can still decipher the mathematical relationship from the provided scattering data points. All growth models exhibit an inverted U-shape, peaking during the intermediate period. This phenomenon is consistent with imaging observations by Gilmore, et al.

[65]. For growth in the out-of-plane directions, symbolic regression exclusively selected the polynomial operators to construct the base models. The general model format can be summarized as a combination of up to fourth order polynomials: $g_r(t) = a*t^4 + b*t^3 + c*t^2 + d*t + e$, where $a, b, c, d, e$ are constants. This result suggests a potential model choice for charactering brain out-of-plane growth patterns. Though the accuracy was degraded by data variance, symbolic regression still proves robust in identifying suitable regional growth models for the human brain cortex, as validated in our recent paper [39]. These predicted models will be integrated into ABAQUS via user-defined subroutines to simulate brain folding evolution.

**3.2. Regional growth models accurately simulate folding evolutions patterns**

Figure 8 illustrates the evolution of cortical folding on simulated brain surface, incorporating the regional growth model. The folding patterns at six key simulation moments are displayed and rendered with the displacement magnitude. As displayed, the non-uniform curvature inherent on initial brain surface (29 post-menstrual weeks) was sufficient to initiated bulking at around 0.2 s and generate intrinsic folding modes, which evolved to form distinct gyri and sulci over time. The resulting folding patterns of regional models are visually different. Specifically, regions 4, 11, and 16 exhibited denser folds compared to regions 1 and 9, indicating shorter folding wavelengths in these regions. This disparity is primarily attributed to differences in initial geometry and potentially the effects of growth. Additionally, for each model, areas with higher initial altitude (characterized in the out-of-plane direction) tend to experience greater displacements, as depicted in dark red areas. Since we applied consistent growth ratios both in tangential and radial directions across each model, the non-uniform displacements likely stem from the in-plane compressions caused by tangential growth under boundary confinements. This suggests that tangential growth has a more significant influence on folding evolutions than radial growth, which is also evident in the larger values observed in tangential growth models compared to radial growth model in Figures 6 and 7.

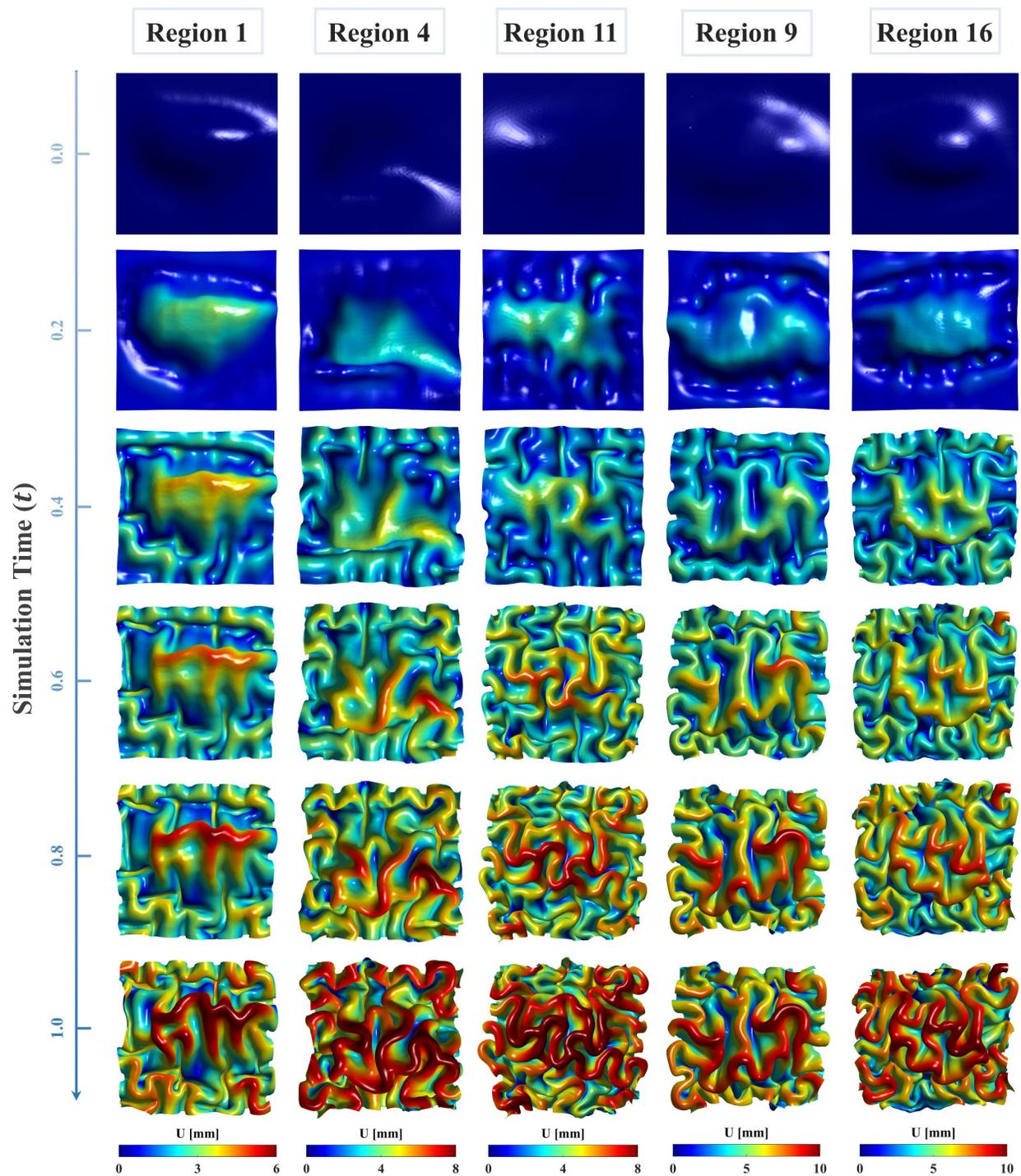

**Figure 8. Longitudinal brain developing patterns for five regions**. Cortical folding patterns of six moments ranging from $t = 0$ to $t = 1.0$ were recorded and rendered with displacement magnitude.

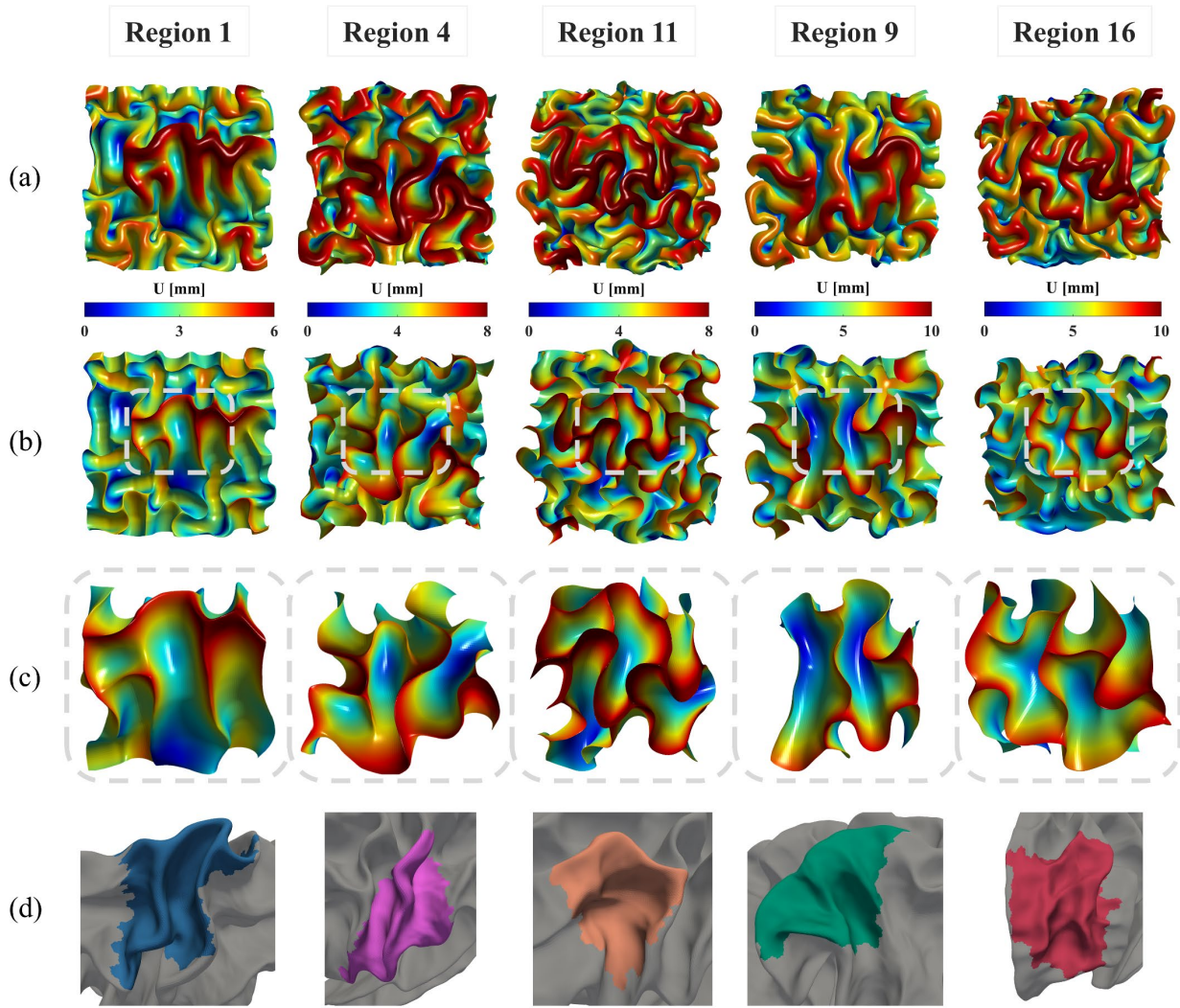

**Figure 9. Simulated folding patterns vs realistic brain images**. (a). Cortical folding patterns of five regions displayed on the cortical surface and rendered with displacement magnitude at $t$ = 1s; (b). Cortical folding patterns of five regions displayed on the gray/white matter interface; (c). Zoom-in representation of the highlighted regions; (d) Five regions from anatomically realistic human brain images.

The nonuniform distribution of displacement remains prominent at the interface between the gray matter and white matter. As shown in Figures 9a and 9b, greater displacements occur in areas with higher initial altitude. Additionally, the simulated folding surfaces exhibit characteristic pattern of cusped sulci and smooth gyri, consistent with the morphology observed in real human brain structures [66]. To validate our simulation results, we focused on the folding patterns at the interface where the initial locations approximately cover the extracted brain regions, as illustrated in Figure 9c. We then compared these patterns with those of a 24-postnatal-month brain, as highlighted in Figure 9d [60]. As shown, our simulated brain folding closely replicates the realistic brain pattern, particularly in regions 1 and 4. For the other three regional models, certain characteristic morphologies are comparable to those in real brains, such as the ladder-shaped concaves in region 16 and the sharp ridge along with a narrow valley in region 9. To quantitatively compare these folding patterns, we further computed the curvature (MC), sulcal depth (SulcDepth), and gyrification index (GI) using the methods described in Section 2.4. The distributions of MC and SulcDepth are shown in the contour plot of Figure 10. As depicted, positive curvatures tend to appear on the gyri, while negative on the sulci. This distinction suggests that mean curvature can serve as a metric to deliberately extract gyri and sulci for further investigations [46, 51, 67].

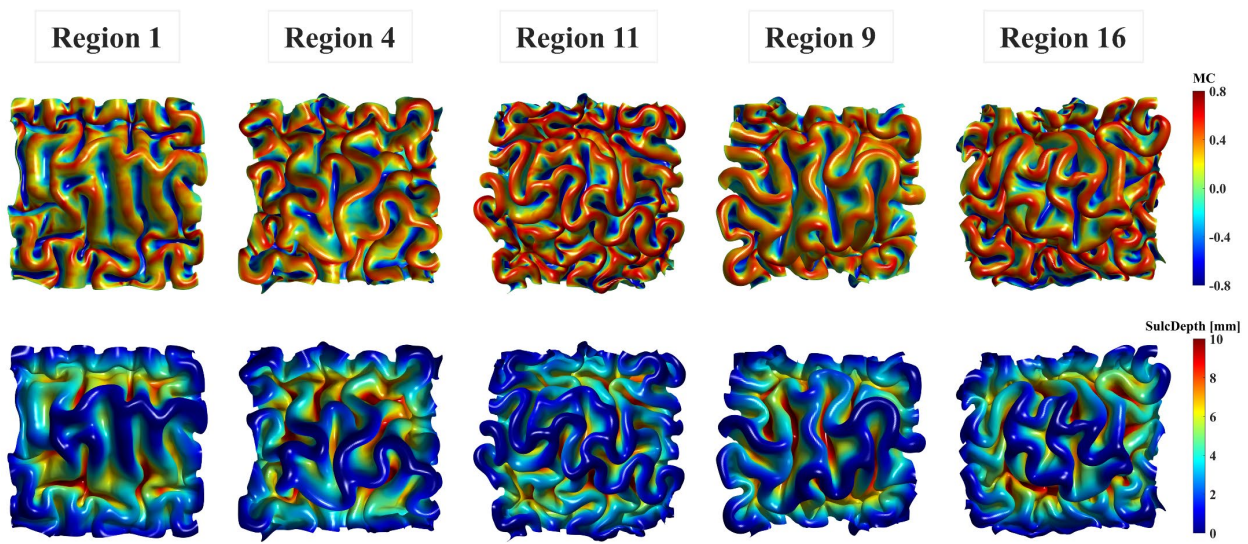

**Figure 10. Mean curvature and sulcal depth**. Distribution of dimensionless mean curvature and sulcal depth of the five selected regions at the final state.

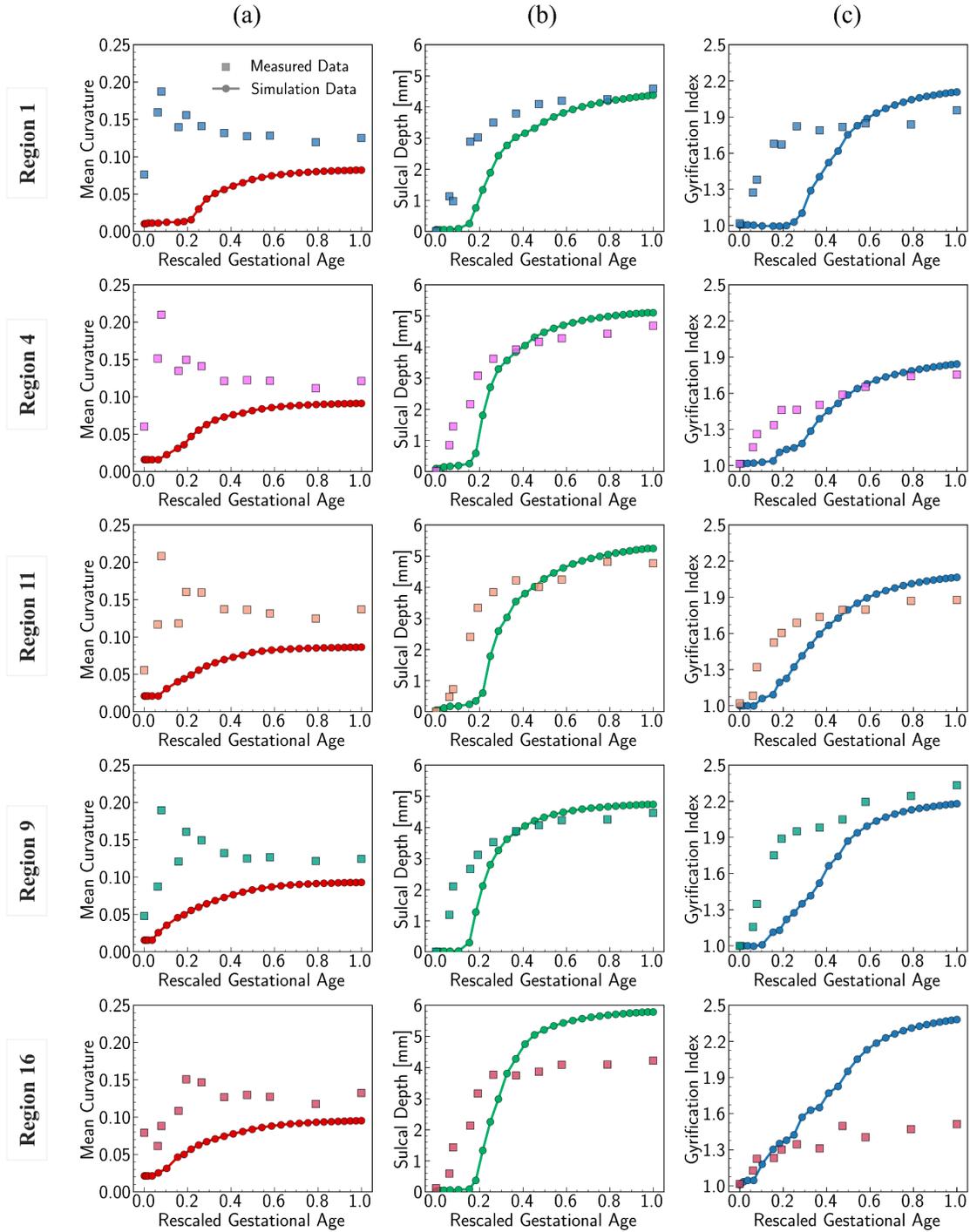

**Figure 11. Quantitative comparison of mean curvature, sulcal depth, and gyrification index among five regions.** Mean curvature (a), sulcal depth (b), and gyrification index (c) of each moment. Ages of the measured data range from 22 post-menstrual weeks to 24 months after birth, and the gestational ages are rescaled into the range of [0, 1] to facilitate data comparison. Mean curvature is determined by averaging absolute dimensionless mean curvature at each point. Squared dots represent data measured from human brain imaging data, while circular dots with lines represent data measured from simulated brain.

Figure 11 shows the comparisons of those quantitative metrics measured in the simulated brain (circular dots and line) with those from a realistic human brain (square dots), with the dots representing the mean values of each metric averaged across the entire ROI surface. As shown in the figure, most of simulation results align well with the realistic brain data in both trend and values. The simulation trajectories exhibit sigmoid-like shapes, starting with flat initial phases, then increasing dramatically after the onset of geometry instabilities (around 0.2 s, as seen in Figure 8), and finally stabilizing once the folding patterns mature. This trend aligns with the identified regional growth models depicted in Figures 6 and 7. Among the three metrics, mean curvature was poorly predicted compared to the other two metrics. It is important to note that the initial curvatures in the simulation are not zero because the simulation models were constructed based on brain regional surfaces, which already display curvatures at the initial state (29 post-menstrual weeks). The sulcal depth and gyrification index correlate well across all regions except for region 16, where a significant deviation occurs after t = 0.4 s. This discrepancy mainly arises from the geometric model being constructed based on a single region. Since region 16 exhibits the highest growth ratio among all selected models, it undergoes the largest deformations during simulation, which, under boundary confinement, would generate the most convoluted folding structures, resulting in the highest GI and sulcal depth. However, in real brain, the compression caused by rapid expansion of region 16 would be relaxed by the surrounding regions, which yields less complex folding patterns such as shallower sulci, as observed in Figure 9. In summary, both qualitative and quantitative comparisons affirm the effectiveness of regional growth models in simulating the brain folding evolution process.

### 3.3. Regional growth model outperforms classic unified growth models

To compare the effectiveness of regional growth models with classic models such as isotropic growth and purely tangential growth, we performed simulations on three regional models with varying growth speeds: the slow-growing region 1, medium-growing region 9, and fast-growing region 16. In both isotropic growth and tangential growth cases, the growth ratio was defined as a constant $g = \sqrt{8}$, which is commonly adopted in previous studies [25, 54, 66]. The performances of different growth theories were compared in Figure 12 (Supplementary Material, Figure S3), where folding patterns at $t = 1$ s were presented and rendered with displacement, mean curvature, and sulcal depth, respectively, to facilitate qualitative comparisons. Quantitative comparisons were also provided by averaging each metric across the entire ROI surface.

As shown in Figures 12a-12c, the regional growth model generates more sulcal and gyral folds compared to the isotropic growth model, especially evident in the intermediate areas corresponding to the initial brain surface. The increased number of folds indicates that regional growth model yields more convoluted folding patterns with shorter wavelengths. This phenomenon concurs with the findings of Budday, et al. [32], which demonstrates that variation in cortical growth modulates secondary instabilities, culminating in highly irregular folding patterns. However, the difference in wavelength between the regional growth model and the tangential growth model is not significant, even though the regional growth model tends to produce more cusped and deep sulci, as well as curled gyri. This suggests that tangential growth theory generates more convoluted patterns than isotropic growth, consistent with previous findings [68, 69]. Furthermore, this validates that tangential (in-plane) growth plays a more dominant role in affecting folding patterns than radial (out-of-plane) growth.

The quantitative comparison illustrated in Figure 12d shows that the regional growth model provides more accurate folding measures than the other two growth theories, with the resulting curvature, sulcal depth, and GI aligning better with real brain data. Additionally, the regional growth model initiates the bulking occurrence earlier and reaches the stabilization period sooner than the other two models. This suggests that the regional growth model can significantly reduce computational costs in simulating brain folding evolution by supplementing a stopping criterion based on stabilization.

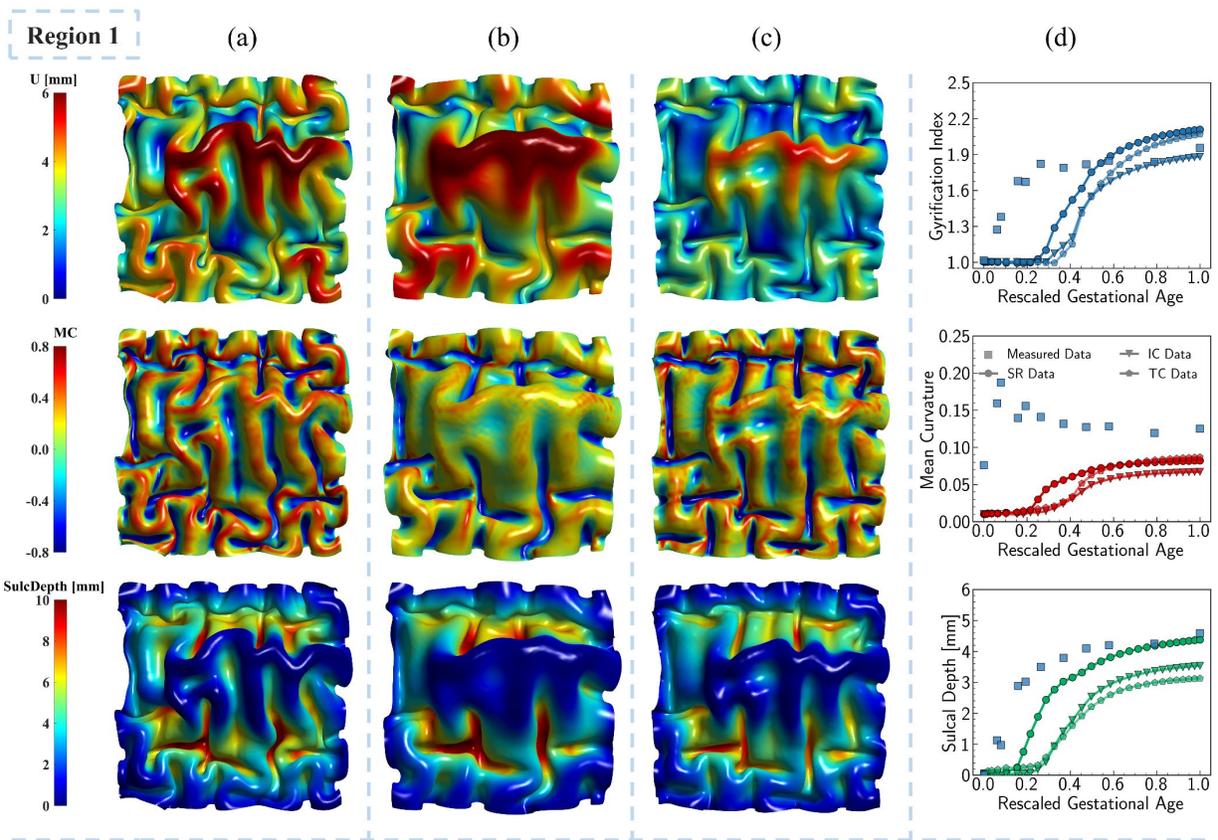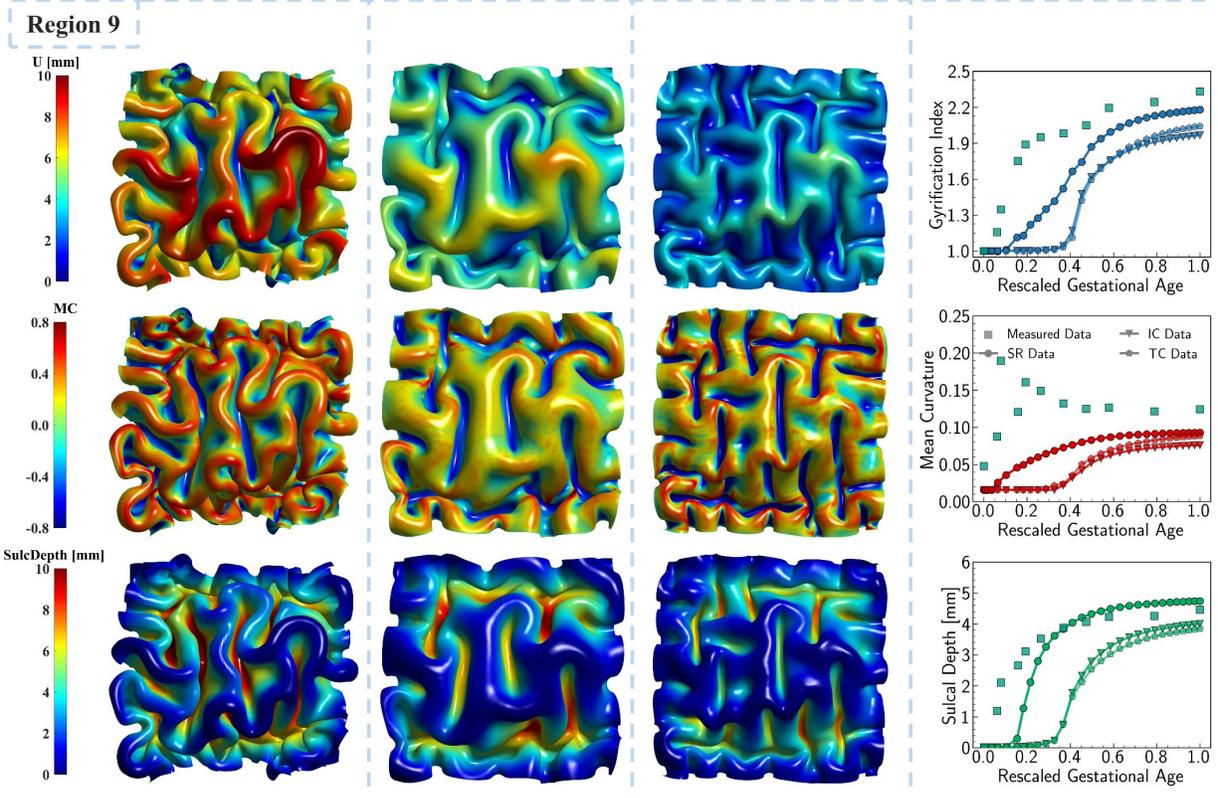

**Figure 12. Symbolic regression growth model vs classic growth model.** Impact of different growth models, including growth model predicted from symbolic regression (SR) algorithm (a), isotropic model with constant growth ratio (IC) model (b), and tangential model with constant growth ratio (TC) (c), on modeling cortical folding patterns. Distributions of displacement (U), mean curvature (MC), and sulcal depth (SulcDepth) of regions 1 and 9 are displayed with their quantitative measure present at the right (d). The growth ratio in both isotropic and tangential growth model is $g = \sqrt{8}$, and no radial growth occurs for the pure tangential model, $g_r = 1$.

### 3.4. Growth ratio values influence folding evolution more than growth trajectory

The regional growth model and purely tangential growth model exhibit similar primary folding patterns, with the regional growth model producing more cusped and deep sulci and curled gyri in secondary details. Both models adopt a similar format of tangential growth, but their values and growth trajectories differ. Noted, the effect of radial growth is ignored due to its minor impact on modulating folding. To figure out whether the value or growth trajectory dominates the folding patterns, we performed simulations on geometric model of region 9 using three growth models with the same resulting growth ratio but different growing trajectories. In addition to the regional growth model, we introduced two additional growth models following linear and Gompertz distributions, respectively, as shown in Figure 13. The Gompertz distribution has been proven successful in modeling tissue growth [70], such as the growth mode of myelinated brain white matter [71] and tumor growth with necrosis [72]. The Gompertz distribution is expressed as follows:

$$g_t(t) = a * \exp\bigl(-\exp(-b * (t - c))\bigr) + 1, \qquad (9)$$

where $a$, $b$, and $c$ are constants. Compared to the regional growth model, the Gompertz model exhibits a typical S shape, starting with an initial flat phase and quickly transitioning into a stabilization phase. In contrast, the linear model has a growth ratio that increases linearly.

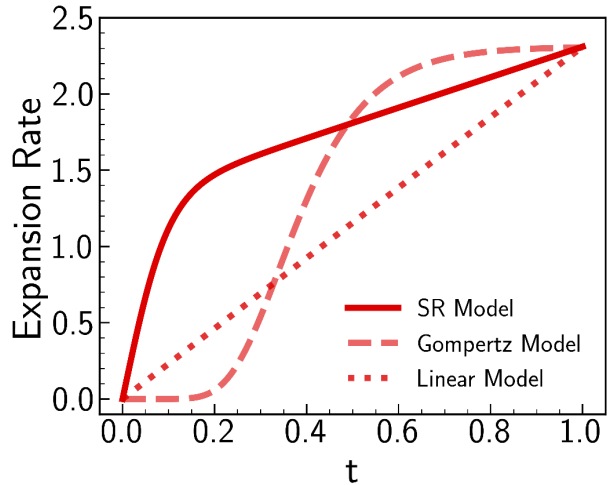

**Figure 13. Three distinct growth models for region 9.** Initial and final value keep the same for all three models, while the developing trajectories are different.

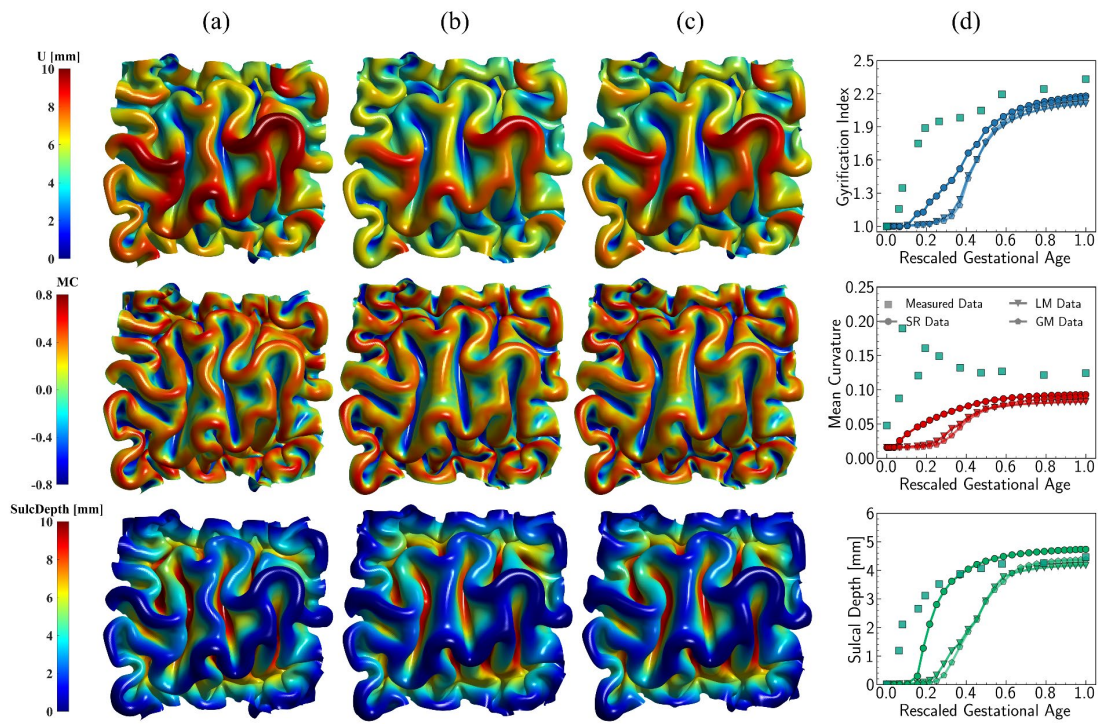

**Figure 14. Impact of growth trajectory on the folding patterns.** Three growth models with distinct developing patterns for region 9, including symbolic regression model (SR) (a), linear growth model (LM) (b), and Gompertz growth model (GM) (c). Distribution of displacement (U), mean curvature (MC), and sulcal depth (SulcDepth) are present with their quantitative measure present at the right (d). In simulations, the above growth models are exclusively functioned in the tangential direction, while the growth in thickness remains consistent, and the radial growth model predicted from symbolic regression is employed, as shown in Figure 6.

Figure 14 illustrates the qualitative comparison of folding patterns rendered with displacement, curvature, and sulcal depth, as well as the quantitative comparison by averaging each metric across the entire ROI surface. As seen in Figures 14a-14c, despite the difference in growing trajectories, the resulting folding demonstrates nearly identical patterns, especially in the linear and Gompertz models. This is also evident in quantitative measurements shown in Figure 14d. Though the regional growth model initiates instabilities earlier due to its rapid growth within the first 0.2 s, the resulting measures of GI and curvature are almost the same across all models. Additionally, the regional growth model generates deeper sulci compared to the other two models. When compared to the findings in Section 3, it becomes apparent that the growth trajectories have a less significant impact on folding patterns than the magnitude of growth itself. This observation remains consistent with the findings of Wang, et al. [50], who report that the cortical growth mode has a limited influence on surface morphology complexity. Interestingly, this result appears to be an intrinsic feature of brain folding, as it persists across various simulation approaches. Our simulations were based on anatomically realistic brain models, incorporating physical accurate growth pattern. In contrast, Wang, et al. [50] employed idealized spherical models with manually defined growth patterns to represent brain structure. Despite these differences in methodology, the core finding—minor impact of growth trajectories on folding patterns—remains consistent.

**3.5. Multi-region model provides more realistic folding results than single-regional model**

All simulations discussed above were conducted on a single brain region, where we applied symmetric confinements to their boundaries to prevent out-of-boundary growth. Despite our efforts to minimize the boundary effects by sufficiently extending the ROI boundaries, this condition may still be overly restrictive in generating compression, which is pivotal for triggering instabilities, resulting in unrealistic folding patterns as observed in region 16 (see Figures 8 and 11). Additionally, in real brains, no region functions or deforms independently without affecting or being affected by surrounding regions. To address these limitations, we constructed a new computational model using the initial geometries of three adjacent regions: regions 1, 4, and 11.

The constructed model is shown in Figure 15a. After proper smoothing and extension, the model uniformly partitions a squared area into three parts, following an approximate proportion of 1:1:2 for regions 1, 11, and 4, respectively.

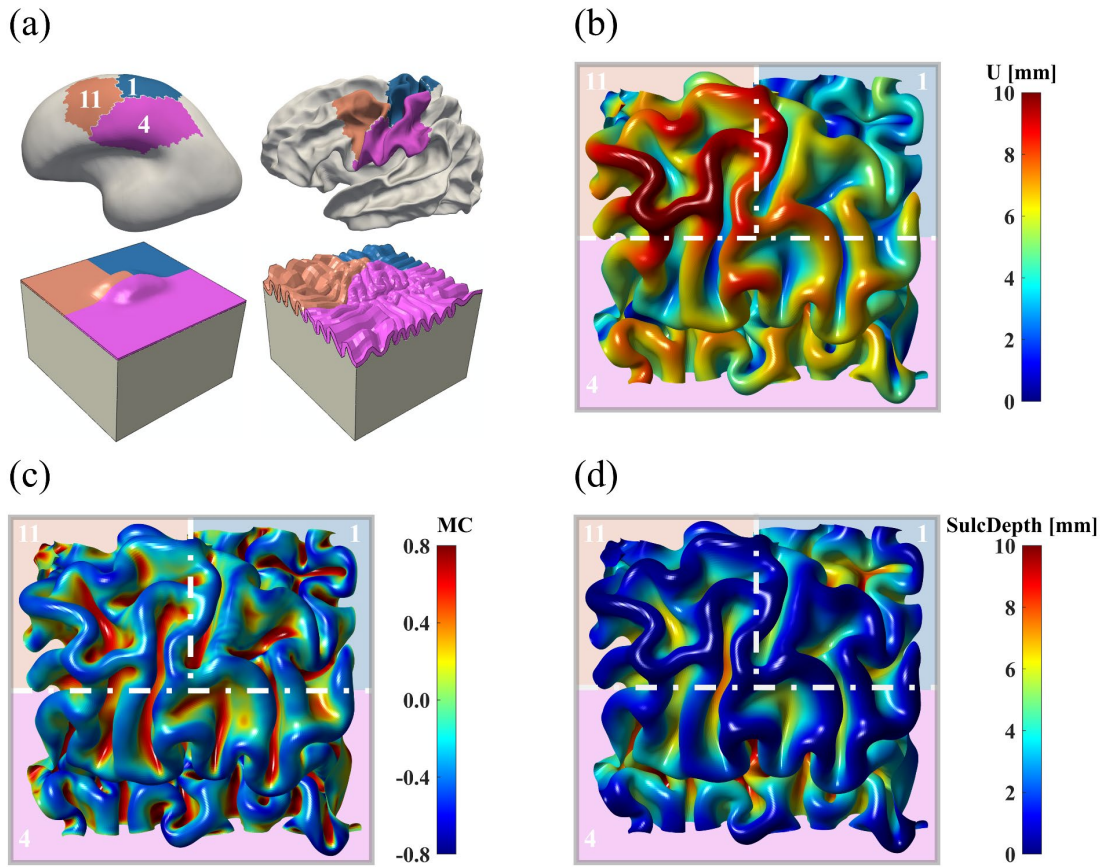

**Figure 15. Brain folding patterns of a multi-region model.** (a). The computational model is based on regions 1, 4, and 11 of the real human brain, with colored areas indicating distinct material properties, specifically growth behaviors. Cortical folding patterns in area of interest are displayed with displacement (b), mean curvature (c), and sulcal depth (d). Regions are delineated by white dotted lines and marked with distinct background colors.

The simulated patterns are illustrated in Figures 15b-15d. Compared to the single-region model (see Figure 8), the multi-region model generates more uniform folds across each region. For example, in the single-region modeling, the folding patterns are fairly condensed in the ROI of region 11 but sparse in region 1. Conversely, these patterns are uniformly distributed in the multi-region model, which is consistent with the anatomical morphology of the real human brain.

However, certain characteristic features like the central sulcus were missed, which could be improved by incorporating more regions into the model construction process. From quantitative comparison in Figure16, it is evident that multi-region model achieves more realistic results compared to the single-region model, with the quantitative measures of GI, mean curvature, and sulcal depth aligning better with the real brain data. This demonstrates the superiority of the multi-region model in simulating brain folding evolutions over the single-region model. In the future, it is expected that a brain-wide model including all 18 regions would provide more promising brain folding both in qualitative patterns and quantitative measures.

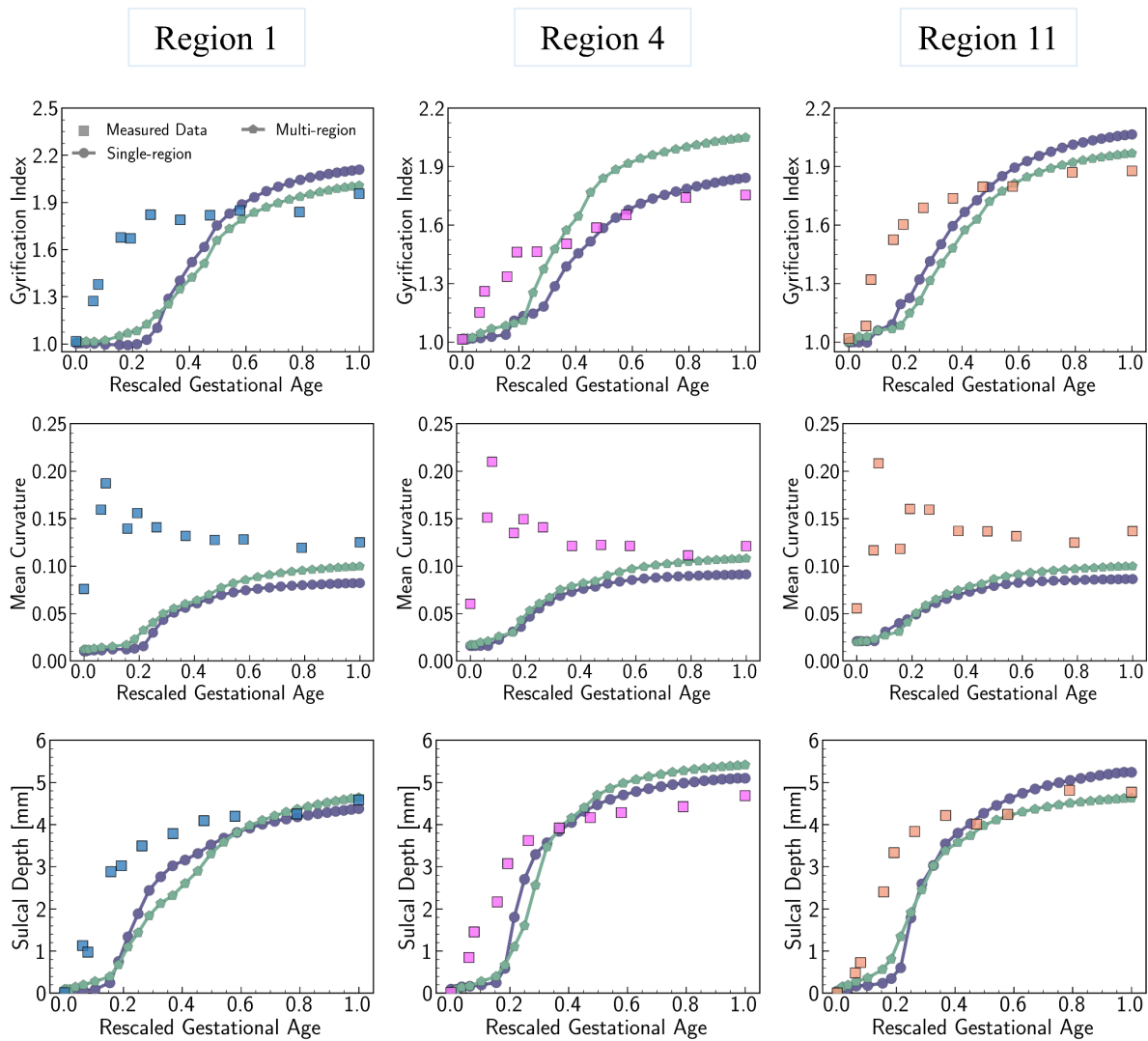

**Figure 16. Multi-region model vs single-region model.** Comparison on the trajectory of gyrification index, mean curvature, and sulcal depth between the multi-region model and single-region model. Mean curvature is determined by averaging absolute dimensionless mean curvature at each point. Squared dots represent data measured from human brain imaging data, while circular dots and pentagonal dots with lines represent data measured from simulation based on single-region model and multi-region model, respectively.

## 4. Discussion

Computational simulations, such as finite element methods, have emerged as promising tools to elucidate the mechanical influence on cortical development, providing valuable insights into the underlying mechanisms [22, 66, 73] and enabling predictive modeling of brain tissue behaviors under various conditions [74-76]. Herein, we show that integrating simulations with regional growth models, derived through symbolic regression using brain developmental data, produces more realistic brain folding patterns. These region-based developing cortical patterns further exhibit accurate alignment with quantitative measures of real brain development. Our findings challenge the conventional views that isotropic, unified growth is sufficient to mimic cortical folding evolutions with modeling results more consistent with the imaging analyses [21, 23]. Instead, our results suggest that heterogenous growth patterns play an indispensable role in modulating brain cortical development, as suggested in the works of Budday, et al. [32], Holland, et al. [77].

The longitudinal brain data encompass surface area expansion measured from 29 post-menstrual weeks to 24 postnatal months [30] and cortical thickness recorded within the first two years after birth [31]. We employed symbolic regression to derive explicit mathematical expressions for growth models across various brain regions, spanning from slow-growing areas to fast-growing areas. The models identified exhibit consistent patterns in describing both in-plane and out-of-plane growth for nearly all regions. This uniformity in regional growth reflects a characteristic growth mode of neural tissues [78] and has also been observed in tumor development [70, 79]. Symbolic regression was chosen over conventional methods, such as multiple regression or neural networks, due to its ability to predict interpretable models from data without requiring prior knowledge of the model structure, which must be predefined in multiple regression [80, 81]. Moreover, symbolic regression explores an infinite functional space to identify candidate models, even with constraints applied to expedite the search process, highlighting its superior performance over other machine-learning-based methods [82, 83], which are restricted to a limited functional space. The explicit mathematical expression of the growth model greatly facilitates the implementation of these models in ABAQUS simulations using user-defined subroutines. Though directly incorporating data into simulation can ensure closer alignments with the provided data, this approach is highly sensitive to noises, particular outliers, and may fail in scenarios requiring further operations, such as differentiation or integration.

Uniform growth theories such as isotropic growth have been extensively employed to model brain development [21-25]. While these methods have significantly advanced our understanding of cortical folding mechanisms, their effectiveness degrades when comparing the resulting cortical folding patterns with imaging observations, particularly in brain-wide simulations [26, 27]. To address this limitation, additional physiological effects such as axonal fibers [23, 77] and cells migrations [84, 85] have been incorporated to more accurately simulate brain folding anatomy. However, these additions inevitably complicate modeling endeavors. Herein, we found that exclusively implementing regional growth into simulations can produce more realistic cortical folding patterns that closely match quantitative measurements of the real brain. This outcome aligns with the principle of Occam's Razor: the fewer deviations from real conditions, the better the simulation's alignment with the observed data. This finding underscores the superiority of heterogeneous growth over homogeneous growth in brain modeling [20]. As presented in Section 3.3, even purely tangential growth can yield more complex folding patterns, similar to those observed in regional growth scenarios, compared to isotropic growth. Introducing such heterogeneity decreases the system's stability by triggering earlier bulking and more convoluted folding patterns. In addition to the regional growth, heterogeneity also exists in other brain properties like stiffness [32], surface curvature [86] and cortical thickness [87]. These factors collectively contribute to the complex evolution of brain folding patterns.

While the brain surface bulking is predominantly driven by intrinsic properties such as initial surface curvature, relative stiffness or thickness ratio between gray matter and white matter, the growth does influence the cortical folding, especially in terms of cortical depth and folding complexity, as revealed by quantitative measures. Moreover, the value of growth ratio plays a more crucial role in modulating folding evolutions compared to the growth trajectory. Intuitively, the external forces, such as the thermal expansion introduced here, do not dictate the bulking modes but rather influence the timing of bulking initialization and the extent of bulking development. However, brain folding is a dynamic process, with numerous complex factors interacting in a coupled manner. Heterogenous growth can alter these intrinsic features. For example, differential growth in the in-plane and out-of-plane directions can change the thickness ratios between the gray matter and white matter. Similarly, axonal maturation, influenced by cell migration, can affect tissues stiffness [88]. These features evolve dynamically, introducing variability into the cortical folding process.

Conducting simulations on a single brain region allows for comparative analysis of distinct growth theories and their effects on brain folding. However, this approach may lead to unrealistic folding patterns due to the artificially imposed boundary conditions. Our findings indicate that a multi-region computational model, which considers three adjacent regions simultaneously, offers a more reliable result by producing more uniformly distributed folding patterns. In the future, a brain-wide model encompassing all 18 parcellated regions is expected to yield more realistic folding predictions, by integrating regional growth models derived through symbolic regression. Moreover, our study can be improved by addressing the following issues: First, we assumed uniform cortical thickness across each brain region. Incorporating anatomically accurate cortical thickness, accounting for both gray and white matter in the model construction process, would provide a more convincing geometric model. Second, the tangential growth within the cortical layer was assumed to be uniform. In reality, this growth varies spatially, as evident in differential growth within the six-layered cortex [89]. Future studies should consider adopting a spatially dependent growth profile, as proposed by Tallinen, et al. [8]. Last but not least, the brain tissue in our model was treated as an incompressible hyperelastic material described by the neo-Hookean strain energy function. Incorporating a regional hyperelastic model with a degree of compressibility, characterized though symbolic regression, could account for the heterogeneity in stiffness and further enhance the reliability of our simulation results.

## 5. Conclusion

In this study, we investigated the impact of regional growth on the computational simulation of brain cortical folding. Using symbolic regression, we derived explicit mathematical growth model using the longitudinal data of cortical surface area expansion and cortical thickness measured on real human brains. These regressed models were then integrated into ABAQUS to simulate the evolution process of cortical folding. The resulting folding patterns were recorded and quantified using several mechanical descriptors, including mean curvature, sulcal depth, and gyrification index. Our findings demonstrate that regional growth models can generate complex folding patterns that closely resemble the anatomical structure of the brain in both quantitative and qualitative aspects. These models outperformed conventional uniform growth models such as isotropic growth and purely tangential growth in accurately modeling brain folding developments. We observed that the magnitude of growth, rather than its trajectory, plays a more predominate role in shaping folding patterns. Moreover, multi-region modeling produces more uniform folding patterns similar to imaging observations compared to single-region models. Our results underscore the importance of implementing regional growth in brain folding simulations. This approach holds promise for advancing early diagnostics of cortical malformations like pachygyria, lissencephaly, and polymicrogyria and improves treatment for neurodevelopmental disorders such as autism and epilepsy.

**Acknowledgement**

JH and XW acknowledges the support from National Science Foundation (IIS-2011369) and National Institutes of Health (1R01NS135574-01). GL is supported in part by National Institutes

of Health (MH116225, MH123202, ES033518, AG075582, NS128534, and NS135574). This work also utilizes approaches developed by an NIH grant (1U01MH110274) and the efforts of the UNC/UMN Baby Connectome Project Consortium. Some of the data was provided by the developing Human Connectome Project, KCL-Imperial-Oxford Consortium funded by the European Research Council under the European Union Seventh Framework Programme (FP/2007-2013) / ERC Grant Agreement no. [319456]. We are grateful to the families who generously supported this research.

**Data Availability Statement**

The original contributions presented in the study are included in the article/supplemental material. Further inquiries can be directed to the corresponding authors. The dHCP dataset is publicly available at the Developing Human Connectome Project repository: http://www.developingconnectome.org [90]. The BCP dataset is publicly available in NIMH Data Archive: https://nda.nih.gov/edit_collection.html?id=2848 [91].

**Conflict of Interest**

The authors declare that the research was conducted in the absence of any commercial or financial relationships that could be construed as a potential conflict of interest.

**Authorship Contribution Statement**

JH: Methodology, Software, Validation, Investigation, Writing – Original Draft; ZW: Data Curation, Writing – Review and Editing; XC: Formal Analysis, Writing – Review and Editing; LW: Validation, Writing – Review and Editing; DZ: Validation, Writing – Review and Editing; TL: Validation, Writing – Review and Editing; GL: Conceptualization, Validation, Writing – Review and Editing; XW: Conceptualization, Validation, Supervision, Funding acquisition, Writing – Original Draft, Writing – Review and Editing.

# Reference

[1] G. Lohmann, D.Y. von Cramon, A.C.F. Colchester, Deep Sulcal Landmarks Provide an Organizing Framework for Human Cortical Folding, Cerebral Cortex, 18 (2007) 1415-1420. http://doi.org/10.1093/cercor/bhm174.

[2] J.E. Schmitt, A. Raznahan, S. Liu, M.C. Neale, The heritability of cortical folding: evidence from the human connectome project, Cerebral Cortex, 31 (2021) 702-715.

[3] J. Zeidan, E. Fombonne, J. Scorah, A. Ibrahim, M.S. Durkin, S. Saxena, A. Yusuf, A. Shih, M. Elsabbagh, Global prevalence of autism: A systematic review update, Autism research, 15 (2022) 778-790.

[4] E. Beghi, The epidemiology of epilepsy, Neuroepidemiology, 54 (2020) 185-191.

[5] R.A. McCutcheon, T.R. Marques, O.D. Howes, Schizophrenia—an overview, JAMA psychiatry, 77 (2020) 201-210.

[6] L. Subramanian, M.E. Calcagnotto, M.F. Paredes, Cortical malformations: lessons in human brain development, Frontiers in cellular neuroscience, 13 (2020) 576.

[7] D.C. Van Essen, Biomechanical models and mechanisms of cellular morphogenesis and cerebral cortical expansion and folding, Seminars in Cell & Developmental Biology, 140 (2023) 90-104. http://doi.org/10.1016/j.semcdb.2022.06.007.

[8] T. Tallinen, J.Y. Chung, J.S. Biggins, L. Mahadevan, Gyrification from constrained cortical expansion, Proceedings of the National Academy of Sciences, 111 (2014) 12667-12672.

[9] E. Takahashi, R.D. Folkerth, A.M. Galaburda, P.E. Grant, Emerging Cerebral Connectivity in the Human Fetal Brain: An MR Tractography Study, Cerebral Cortex, 22 (2011) 455-464. http://doi.org/10.1093/cercor/bhr126.

[10] R.M. Fame, M.K. Lehtinen, Emergence and developmental roles of the cerebrospinal fluid system, Developmental cell, 52 (2020) 261-275.

[11] L.F. Franchini, Genetic Mechanisms Underlying Cortical Evolution in Mammals, Frontiers in Cell and Developmental Biology, 9 (2021). http://doi.org/10.3389/fcell.2021.591017.

[12] P.V. Bayly, L.A. Taber, C.D. Kroenke, Mechanical forces in cerebral cortical folding: a review of measurements and models, J Mech Behav Biomed Mater, 29 (2014) 568-581. http://doi.org/10.1016/j.jmbbm.2013.02.018.

[13] C.D. Kroenke, P.V. Bayly, How forces fold the cerebral cortex, Journal of Neuroscience, 38 (2018) 767-775.

[14] D.C.V. Essen, A tension-based theory of morphogenesis and compact wiring in the central nervous system, Nature, 385 (1997) 313-318. http://doi.org/10.1038/385313a0.

[15] J. Nie, L. Guo, K. Li, Y. Wang, G. Chen, L. Li, H. Chen, F. Deng, X. Jiang, T. Zhang, L. Huang, C. Faraco, D. Zhang, C. Guo, P.-T. Yap, X. Hu, G. Li, J. Lv, Y. Yuan, D. Zhu, J. Han, D. Sabatinelli, Q. Zhao, L.S. Miller, B. Xu, P. Shen, S. Platt, D. Shen, X. Hu, T. Liu, Axonal Fiber Terminations Concentrate on Gyri, Cerebral Cortex, 22 (2011) 2831-2839. http://doi.org/10.1093/cercor/bhr361.

[16] G. Xu, A.K. Knutsen, K. Dikranian, C.D. Kroenke, P.V. Bayly, L.A. Taber, Axons pull on the brain, but tension does not drive cortical folding, J Biomech Eng, 132 (2010) 071013-071013.

[17] L. Ronan, N. Voets, C. Rua, A. Alexander-Bloch, M. Hough, C. Mackay, T.J. Crow, A. James, J.N. Giedd, P.C. Fletcher, Differential tangential expansion as a mechanism for cortical gyrification, Cerebral Cortex, 24 (2014) 2219-2228. http://doi.org/10.1093/cercor/bht082.

[18] M.R. Rosenzweig, Effects of differential experience on the brain and behavior, Developmental neuropsychology, 24 (2003) 523-540.

[19] D.P. Richman, R.M. Stewart, J. Hutchinson, V.S. Caviness Jr, Mechanical Model of Brain Convolutional Development, Science, 189 (1975) 18-21. http://doi.org/10.1126/science.113562.

[20] M. Darayi, M.E. Hoffman, J. Sayut, S. Wang, N. Demirci, J. Consolini, M.A. Holland, Computational models of cortical folding: A review of common approaches, Journal of Biomechanics, 139 (2022) 110851. http://doi.org/10.1016/j.jbiomech.2021.110851.

[21] P.V. Bayly, R.J. Okamoto, G. Xu, Y. Shi, L.A. Taber, A cortical folding model incorporating stress-dependent growth explains gyral wavelengths and stress patterns in the developing brain, Physical Biology, 10 (2013) 016005. http://doi.org/10.1088/1478-3975/10/1/016005.

[22] S. Budday, P. Steinmann, E. Kuhl, The role of mechanics during brain development, Journal of the Mechanics and Physics of Solids, 72 (2014) 75-94.

[23] P. Chavoshnejad, X. Li, S. Zhang, W. Dai, L. Vasung, T. Liu, T. Zhang, X. Wang, M.J. Razavi, Role of axonal fibers in the cortical folding patterns: A tale of variability and regularity, Brain Multiphysics, 2 (2021) 100029. http://doi.org/10.1016/j.brain.2021.100029.

[24] M. Holland, S. Budday, A. Goriely, E. Kuhl, Symmetry breaking in wrinkling patterns: Gyri are universally thicker than sulci, Physical review letters, 121 (2018) 228002.

[25] M.J. Razavi, T. Liu, X. Wang, Mechanism Exploration of 3-Hinge Gyral Formation and Pattern Recognition, Cerebral Cortex Communications, 2 (2021). http://doi.org/10.1093/texcom/tgab044.

[26] J. Nie, L. Guo, G. Li, C. Faraco, L. Stephen Miller, T. Liu, A computational model of cerebral cortex folding, Journal of Theoretical Biology, 264 (2010) 467-478. http://doi.org/10.1016/j.jtbi.2010.02.002.

[27] S.N. Verner, K. Garikipati, A computational study of the mechanisms of growth-driven folding patterns on shells, with application to the developing brain, Extreme Mechanics Letters, 18 (2018) 58-69. http://doi.org/10.1016/j.eml.2017.11.003.

[28] K.E. Garcia, X. Wang, S.E. Santiago, S. Bakshi, A.P. Barnes, C.D. Kroenke, Longitudinal MRI of the developing ferret brain reveals regional variations in timing and rate of growth, Cerebral Cortex, 34 (2024). http://doi.org/10.1093/cercor/bhae172.

[29] C.D. Kroenke, E.N. Taber, L.A. Leigland, A.K. Knutsen, P.V. Bayly, Regional patterns of cerebral cortical differentiation determined by diffusion tensor MRI, Cerebral Cortex, 19 (2009) 2916-2929.

[30] Y. Huang, Z. Wu, F. Wang, D. Hu, T. Li, L. Guo, L. Wang, W. Lin, G. Li, Mapping developmental regionalization and patterns of cortical surface area from 29 post-menstrual weeks to 2 years of age, Proceedings of the National Academy of Sciences, 119 (2022) e2121748119. http://doi.org/10.1073/pnas.2121748119.

[31] F. Wang, C. Lian, Z. Wu, H. Zhang, T. Li, Y. Meng, L. Wang, W. Lin, D. Shen, G. Li, Developmental topography of cortical thickness during infancy, Proceedings of the National Academy of Sciences, 116 (2019) 15855-15860. http://doi.org/10.1073/pnas.1821523116.

[32] S. Budday, P. Steinmann, On the influence of inhomogeneous stiffness and growth on mechanical instabilities in the developing brain, International Journal of Solids and Structures, 132 (2018) 31-41. http://doi.org/10.1016/j.ijsolstr.2017.08.010.

[33] S. Wang, N. Demirci, M.A. Holland, Numerical investigation of biomechanically coupled growth in cortical folding, Biomechanics and Modeling in Mechanobiology, 20 (2021) 555-567.

[34] T. Zhang, M.J. Razavi, X. Li, H.B. Chen, T.M. Liu, X.Q. Wang, Mechanism of Consistent Gyrus Formation: an Experimental and Computational Study, Scientific Reports, 6 (2016). http://doi.org/10.1038/srep37272.

[35] E.K. Rodriguez, A. Hoger, A.D. McCulloch, Stress-dependent finite growth in soft elastic tissues, Journal of biomechanics, 27 (1994) 455-467. http://doi.org/10.1016/0021-9290(94)90021-3.

[36] G.A. Holzapfel, Nonlinear solid mechanics: a continuum approach for engineering science, Kluwer Academic Publishers Dordrecht, 2002.

[37] D. Angelis, F. Sofos, T.E. Karakasidis, Artificial Intelligence in Physical Sciences: Symbolic Regression Trends and Perspectives, Archives of Computational Methods in Engineering, 30 (2023) 3845-3865. http://doi.org/10.1007/s11831-023-09922-z.

[38] B. Bahmani, W. Sun, Physics‐constrained symbolic model discovery for polyconvex incompressible hyperelastic materials, International Journal for Numerical Methods in Engineering, 125 (2024) e7473.

[39] J. Hou, X. Chen, T. Wu, E. Kuhl, X. Wang, Automated Data-Driven Discovery of Material Models Based on Symbolic Regression: A Case Study on Human Brain Cortex, arXiv preprint arXiv:2402.05238, (2024).

[40] Z. Zhang, Z. Zou, E. Kuhl, G.E. Karniadakis, Discovering a reaction–diffusion model for Alzheimer's disease by combining PINNs with symbolic regression, Computer Methods in Applied Mechanics and Engineering, 419 (2024) 116647.

[41] W. Ben Chaabene, M.L. Nehdi, Genetic programming based symbolic regression for shear capacity prediction of SFRC beams, Construction and Building Materials, 280 (2021) 122523. http://doi.org/10.1016/j.conbuildmat.2021.122523.

[42] M. Cranmer, Interpretable machine learning for science with PySR and SymbolicRegression.jl, arXiv preprint arXiv:2305.01582, (2023).

[43] J. Dubois, M. Alison, S.J. Counsell, L. Hertz‐Pannier, P.S. Hüppi, M.J. Benders, MRI of the neonatal brain: a review of methodological challenges and neuroscientific advances, Journal of Magnetic Resonance Imaging, 53 (2021) 1318-1343.

[44] F. Ge, X. Li, M.J. Razavi, H. Chen, T. Zhang, S. Zhang, L. Guo, X. Hu, X. Wang, T. Liu, Denser Growing Fiber Connections Induce 3-hinge Gyral Folding, Cerebral Cortex, 28 (2018) 1064-1075. http://doi.org/10.1093/cercor/bhx227.

[45] G. Abaqus, Abaqus 6.11, Dassault Systemes Simulia Corporation, Providence, RI, USA, 3 (2011).

[46] P. Chavoshnejad, L. Chen, X. Yu, J. Hou, N. Filla, D. Zhu, T. Liu, G. Li, M.J. Razavi, X. Wang, An integrated finite element method and machine learning algorithm for brain morphology prediction, Cerebral Cortex, 33 (2023) 9354-9366. http://doi.org/10.1093/cercor/bhad208.

[47] M.R.T.L.X.W. Mir Jalil Razavi, Mechanical role of a growing solid tumor on cortical folding, Computer Methods in Biomechanics and Biomedical Engineering, (2017).

[48] J. Weickenmeier, R. de Rooij, S. Budday, T.C. Ovaert, E. Kuhl, The mechanical importance of myelination in the central nervous system, Journal of the Mechanical Behavior of Biomedical Materials, 76 (2017) 119-124. http://doi.org/10.1016/j.jmbbm.2017.04.017.
[49] Y. Cao, Y. Jiang, B. Li, X. Feng, Biomechanical modeling of surface wrinkling of soft tissues with growth-dependent mechanical properties, Acta Mechanica Solida Sinica, 25 (2012) 483-492. http://doi.org/10.1016/S0894-9166(12)60043-3.
[50] X. Wang, J. Lefèvre, A. Bohi, M.A. Harrach, M. Dinomais, F. Rousseau, The influence of biophysical parameters in a biomechanical model of cortical folding patterns, Scientific Reports, 11 (2021) 7686. http://doi.org/10.1038/s41598-021-87124-y.
[51] S. Zhang, P. Chavoshnejad, X. Li, L. Guo, X. Jiang, J. Han, L. Wang, G. Li, X. Wang, T. Liu, M.J. Razavi, S. Zhang, T. Zhang, Gyral peaks: Novel gyral landmarks in developing macaque brains, Human Brain Mapping, 43 (2022) 4540-4555. http://doi.org/10.1002/hbm.25971.
[52] D. Duan, S. Xia, I. Rekik, Z. Wu, L. Wang, W. Lin, J.H. Gilmore, D. Shen, G. Li, Individual identification and individual variability analysis based on cortical folding features in developing infant singletons and twins, Human Brain Mapping, 41 (2020) 1985-2003. http://doi.org/10.1002/hbm.24924.
[53] M. Meyer, M. Desbrun, P. Schröder, A.H. Barr, Discrete Differential-Geometry Operators for Triangulated 2-Manifolds, in: H.-C. Hege, K. Polthier (Eds.) Visualization and Mathematics III, Springer Berlin Heidelberg, Berlin, Heidelberg, 2003, pp. 35-57.
[54] R. Balouchzadeh, P.V. Bayly, K.E. Garcia, Effects of stress-dependent growth on evolution of sulcal direction and curvature in models of cortical folding, Brain Multiphysics, 4 (2023) 100065. http://doi.org/10.1016/j.brain.2023.100065.
[55] H.J. Yun, K. Im, J.-J. Yang, U. Yoon, J.-M. Lee, Automated sulcal depth measurement on cortical surface reflecting geometrical properties of sulci, PloS one, 8 (2013) e55977.
[56] A. Makropoulos, E.C. Robinson, A. Schuh, R. Wright, S. Fitzgibbon, J. Bozek, S.J. Counsell, J. Steinweg, K. Vecchiato, J. Passerat-Palmbach, The developing human connectome project: A minimal processing pipeline for neonatal cortical surface reconstruction, Neuroimage, 173 (2018) 88-112.
[57] E.J. Hughes, T. Winchman, F. Padormo, R. Teixeira, J. Wurie, M. Sharma, M. Fox, J. Hutter, L. Cordero‐Grande, A.N. Price, A dedicated neonatal brain imaging system, Magnetic resonance in medicine, 78 (2017) 794-804.
[58] L. Zubiaurre-Elorza, S. Soria-Pastor, C. Junque, R. Sala-Llonch, D. Segarra, N. Bargallo, A. Macaya, Cortical Thickness and Behavior Abnormalities in Children Born Preterm, PLOS ONE, 7 (2012) e42148. http://doi.org/10.1371/journal.pone.0042148.
[59] G. Li, L. Wang, F. Shi, J.H. Gilmore, W. Lin, D. Shen, Construction of 4D high-definition cortical surface atlases of infants: Methods and applications, Medical Image Analysis, 25 (2015) 22-36. http://doi.org/10.1016/j.media.2015.04.005.
[60] Z. Wu, L. Wang, W. Lin, J.H. Gilmore, G. Li, D. Shen, Construction of 4D infant cortical surface atlases with sharp folding patterns via spherical patch-based group-wise sparse representation, Human Brain Mapping, 40 (2019) 3860-3880. http://doi.org/10.1002/hbm.24636.
[61] A.E. Lyall, F. Shi, X. Geng, S. Woolson, G. Li, L. Wang, R.M. Hamer, D. Shen, J.H. Gilmore, Dynamic Development of Regional Cortical Thickness and Surface Area in Early Childhood, Cerebral Cortex, 25 (2014) 2204-2212. http://doi.org/10.1093/cercor/bhu027.

[62] H. Kalantar-Hormozi, R. Patel, A. Dai, J. Ziolkowski, H.-M. Dong, A. Holmes, A. Raznahan, G.A. Devenyi, M.M. Chakravarty, A cross-sectional and longitudinal study of human brain development: The integration of cortical thickness, surface area, gyrification index, and cortical curvature into a unified analytical framework, NeuroImage, 268 (2023) 119885. http://doi.org/10.1016/j.neuroimage.2023.119885.

[63] H. Huang, R. Xue, J. Zhang, T. Ren, L.J. Richards, P. Yarowsky, M.I. Miller, S. Mori, Anatomical characterization of human fetal brain development with diffusion tensor magnetic resonance imaging, Journal of Neuroscience, 29 (2009) 4263-4273. http://doi.org/10.1523/JNEUROSCI.2769-08.2009.

[64] M. Ouyang, T. Jeon, A. Sotiras, Q. Peng, V. Mishra, C. Halovanic, M. Chen, L. Chalak, N. Rollins, T.P.L. Roberts, C. Davatzikos, H. Huang, Differential cortical microstructural maturation in the preterm human brain with diffusion kurtosis and tensor imaging, Proceedings of the National Academy of Sciences, 116 (2019) 4681-4688. http://doi.org/10.1073/pnas.1812156116.

[65] J.H. Gilmore, R.C. Knickmeyer, W. Gao, Imaging structural and functional brain development in early childhood, Nature Reviews Neuroscience, 19 (2018) 123-137.

[66] T. Tallinen, J.Y. Chung, F. Rousseau, N. Girard, J. Lefèvre, L. Mahadevan, On the growth and form of cortical convolutions, Nature Physics, 12 (2016) 588-593. http://doi.org/10.1038/nphys3632.

[67] T. Zhang, H. Chen, M.J. Razavi, Y. Li, F. Ge, L. Guo, X. Wang, T. Liu, Exploring 3-hinge gyral folding patterns among HCP Q3 868 human subjects, Human Brain Mapping, 39 (2018) 4134-4149. http://doi.org/10.1002/hbm.24237.

[68] P. Chavoshnejad, L. Vallejo, S. Zhang, Y. Guo, W. Dai, T. Zhang, M.J. Razavi, Mechanical hierarchy in the formation and modulation of cortical folding patterns, Scientific Reports, 13 (2023) 13177. http://doi.org/10.1038/s41598-023-40086-9.

[69] K.E. Garcia, C.D. Kroenke, P.V. Bayly, Mechanics of cortical folding: stress, growth and stability, Philosophical Transactions of the Royal Society B-Biological Sciences, 373 (2018). http://doi.org/10.1098/rstb.2017.0321.

[70] M. Peterson, B.C. Warf, S.J. Schiff, Normative human brain volume growth, J Neurosurg Pediatr, 21 (2018) 478-485. http://doi.org/10.3171/2017.10.Peds17141.

[71] N. Sadeghi, M. Prastawa, P.T. Fletcher, J. Wolff, J.H. Gilmore, G. Gerig, Regional characterization of longitudinal DT-MRI to study white matter maturation of the early developing brain, Neuroimage, 68 (2013) 236-247.

[72] C. Vaghi, A. Rodallec, R. Fanciullino, J. Ciccolini, J.P. Mochel, M. Mastri, C. Poignard, J.M. Ebos, S. Benzekry, Population modeling of tumor growth curves and the reduced Gompertz model improve prediction of the age of experimental tumors, PLoS computational biology, 16 (2020) e1007178.

[73] M. Jalil Razavi, T. Zhang, T. Liu, X. Wang, Cortical Folding Pattern and its Consistency Induced by Biological Growth, Scientific Reports, 5 (2015) 14477. http://doi.org/10.1038/srep14477.

[74] M. Alenyà, X. Wang, J. Lefèvre, G. Auzias, B. Fouquet, E. Eixarch, F. Rousseau, O. Camara, Computational pipeline for the generation and validation of patient-specific mechanical models of brain development, Brain Multiphysics, 3 (2022). http://doi.org/10.1016/j.brain.2022.100045.

[75] K.E. Garcia, X. Wang, C.D. Kroenke, A model of tension-induced fiber growth predicts white matter organization during brain folding, Nature Communications, 12 (2021) 6681. http://doi.org/10.1038/s41467-021-26971-9.

[76] M.A. Holland, S. Budday, G. Li, D. Shen, A. Goriely, E. Kuhl, Folding drives cortical thickness variations, The European Physical Journal Special Topics, 229 (2020) 2757-2778. http://doi.org/10.1140/epjst/e2020-000001-6.
[77] M.A. Holland, K.E. Miller, E. Kuhl, Emerging brain morphologies from axonal elongation, Annals of biomedical engineering, 43 (2015) 1640-1653. http://doi.org/10.1007/s10439-015-1312-9.
[78] E.M.W. Billig, W.P. O'Meara, E.M. Riley, F.E. McKenzie, Developmental allometry and paediatric malaria, Malaria Journal, 11 (2012) 64. http://doi.org/10.1186/1475-2875-11-64.
[79] K.R. Swanson, C. Bridge, J. Murray, E.C. Alvord Jr, Virtual and real brain tumors: using mathematical modeling to quantify glioma growth and invasion, Journal of the neurological sciences, 216 (2003) 1-10.
[80] N. Filla, J. Hou, T. Liu, S. Budday, X. Wang, Accuracy meets simplicity: A constitutive model for heterogenous brain tissue, Journal of the Mechanical Behavior of Biomedical Materials, 150 (2024) 106271. http://doi.org/10.1016/j.jmbbm.2023.106271.
[81] J. Hou, N. Filla, X. Chen, M.J. Razavi, T. Liu, X. Wang, Exploring hyperelastic material model discovery for human brain cortex: multivariate analysis vs. artificial neural network approaches, arXiv preprint arXiv:2310.10762, (2023).
[82] M. Flaschel, S. Kumar, L. De Lorenzis, Automated discovery of generalized standard material models with EUCLID, Computer Methods in Applied Mechanics and Engineering, 405 (2023) 115867. http://doi.org/10.1016/j.cma.2022.115867.
[83] K. Linka, S.R.S. Pierre, E. Kuhl, Automated model discovery for human brain using Constitutive Artificial Neural Networks, Acta Biomaterialia, 160 (2023) 134-151.
[84] R.d. Rooij, E. Kuhl, A physical multifield model predicts the development of volume and structure in the human brain, Journal of the Mechanics and Physics of Solids, 112 (2018) 563-576. http://doi.org/10.1016/j.jmps.2017.12.011.
[85] M.S. Zarzor, S. Kaessmair, P. Steinmann, I. Blümcke, S. Budday, A two-field computational model couples cellular brain development with cortical folding, Brain Multiphysics, 2 (2021). http://doi.org/10.1016/j.brain.2021.100025.
[86] S. Budday, P. Steinmann, A. Goriely, E. Kuhl, Size and curvature regulate pattern selection in the mammalian brain, Extreme Mechanics Letters, 4 (2015) 193-198.
[87] N. Demirci, M.A. Holland, Cortical thickness systematically varies with curvature and depth in healthy human brains, Human Brain Mapping, 43 (2022) 2064-2084. http://doi.org/10.1002/hbm.25776.
[88] J. Guo, G. Bertalan, D. Meierhofer, C. Klein, S. Schreyer, B. Steiner, S. Wang, R.V. da Silva, C. Infante-Duarte, S. Koch, Brain maturation is associated with increasing tissue stiffness and decreasing tissue fluidity, Acta biomaterialia, 99 (2019) 433-442.
[89] S. Wang, K. Saito, H. Kawasaki, M.A. Holland, Orchestrated neuronal migration and cortical folding: A computational and experimental study, PLOS Computational Biology, 18 (2022) e1010190. http://doi.org/10.1371/journal.pcbi.1010190.
[90] Developing Human Connectome Project. http://www.developingconnectome.org/project. Accessed 2024.

[91] B.R. Howell, M.A. Styner, W. Gao, P.-T. Yap, L. Wang, K. Baluyot, E. Yacoub, G. Chen, T. Potts, A. Salzwedel, The UNC/UMN Baby Connectome Project (BCP): An overview of the study design and protocol development, NeuroImage, 185 (2019) 891-905.